\documentclass[preprint]{elsarticle}

\usepackage{hyperref,placeins}

\usepackage{amsmath}

\journal{International Journal of Solids and Structures}

\usepackage{numcompress}\biboptions{authoryear}

\def\picHeight{60mm}

\begin{document}

\begin{frontmatter}

\title{Hyperelastic constitutive modeling with exponential decay and application to a viscoelastic adhesive}
\tnotetext[mytitlenote]{\copyright 2018. This accepted manuscript version is made available under the CC-BY-NC-ND 4.0 license http://creativecommons.org/licenses/by-nc-nd/4.0/. The article is published in the International Journal of Solids and Structures,
Volumes 141–-142, 2018, Pages 60--72, ISSN 0020-7683, https://doi.org/10.1016/j.ijsolstr.2018.02.011 .}

\author[ifam]{Olaf Hesebeck\corref{mycorrespondingauthor}}
\cortext[mycorrespondingauthor]{Corresponding author}
\ead{olaf.hesebeck@ifam.fraunhofer.de}

\author[ifam]{Andreas Wulf}
\ead{andreas.wulf@ifam.fraunhofer.de}

\address[ifam]{Fraunhofer Institute for Manufacturing Technology and Advanced Materials, \\
Wiener Stra\ss{}e~12, 28359~Bremen, Germany}

\begin{abstract}
Hyperelastic materials models are well established to describe the non-linear stress-strain relations of elastomers. In this paper, a polyurethane adhesive is considered as an exemplary material and subjected to tensile, compressive and shear tests. Conventional hyperelastic models are unable to capture the mechanical behaviour satisfyingly: If the model agrees with the test results at large strains, it underestimates the stiffness at low strains significantly. We propose a model extension to describe a kind of exponential decay of stiffness which bears some similarity with a specific model developed by Yeoh. The new hyperelastic model is coupled with linear viscoelasticity to account for the strain rate dependence observed for the tested material.

The model parameters are identified using a tensile test at a single strain rate only and a creep test. Furthermore, the compressibility is determined by comparison of the stiffness in two butt joint tests with different aspect rations of the adhesive layer. Using this parameter set the new model provides a very good prediction of tensile, compressive and shear tests at different strain rates.
\end{abstract}


\end{frontmatter}

\section{Introduction}

Elastomers have the capability to undergo very large deformations and exhibit non-linear stress-strain relations. Two different approaches have been developed in the last decades to model their mechanical behaviour: micromechanical network models and hyperelastic potential models.

The first approach considers the molecular chains in the elastomer from a statistical point of view. \citet{kuhn1936beziehungen} calculated the entropy of an idealized molecular chain depending on its stretch and derived the force opposing the elongation of the chain. In the molecular network of an elastomer the chains have arbitrary orientations, so that a kind of averaging is required to obtain the global material response \citep{treloar1954photoelastic,wu1993improved}. A well known example for a model considering a finite number of orientations is the 8-chain model of \citet{arruda1993three}. Recent work on this modeling approach considers chain interactions, the contributions of entropy as well as energy, statistics of chain length distributions, Mullin's effect, and viscoelasticity \citep[see e.g.][]{miehe2004micro,miehe2005micro,goktepe2005micro,balabaev2009extension,itskov2016rubber,lorenz2012mikrostruktur,rebouah2014permanent,smeulders1999phenomenological}. While this class of models offers the chance to simulate complex material behavior using a comparatively low number of physically meaningful parameters, the numerical effort required to realize the network models in a finite element analysis is usually much higher than for phenomenological models.

Phenomenological models adopt a continuum mechanics point of view and disregard the microscopic structure of the material. The hyperelastic material models describe the non-linear elastic behavior by formulating the strain energy density as a function of the deformation state. This elastic potential is expressed as a function of either the strain invariants or the principal stretches. An example for an early hyperelastic model is the work of \citet{mooney1940theory} which formulates the potential of an incompressible material as a function of the invariants $I_1$ and $I_2$ of the right Cauchy-Green strain tensor. \citet{Rivlin379} proposed a generalization to a polynomial of the strain invariants.  While polynomial models of high order can provide a good approximation of a given stress-strain curve, high order models are not well suited for extrapolation and have a limited range of stability. A deliberate restriction of admissible stretches was proposed by \citet{gent1996} to consider the limited extensibility of polymer chains. The resulting model bears similarity to the 8-chain model \citep{boyce1996}. \citet{Ogden565} formulated a hyperelastic potential based on the principal stretches instead of the strain invariants. This potential consists of power functions of the stretches with arbitrary exponents, so that in some cases measured stress-strain curves can be fitted well using a lower number of model parameters than polynomial models need. A similar power law formulation using invariants was suggested by \citet{SWANSON198581}. A large number of hyperelastic models has been developed so far, most of them based on the strain invariants. Reviews and comparisons have been recently published by \citet{marckmann2006,Martins2006,vahapoglu2006,Steinmann2012,Chagnon2015}.

\citet{yeoh1990} noted in investigations on a black-filled rubber that the material showed a significant decrease of shear modulus at low strains which could not be described well using a hyperelastic model formulated in a polynomial of the invariants. At first he suggested to explain this material behavior by filler-filler or filler-rubber interactions, but later he observed the same phenomena in unfilled rubber \citep{yeoh1993}, so that he had to discard this explanation. \citet{yeoh1993} proposed to consider the observation in the hyperelastic model by adding an exponentially decaying term to the strain energy density of the model today known as Yeoh model, a reduced polynomial model of third order. A combination of this exponential term with the model of \citet{gent1996} was developed by \citet{yeoh1997}. On the level of statistical models, \citet{thomas1955} proposed an additional term to the force extension relation of a single molecular chain which leads also to a decrease in stiffness for small strains.

Subject of the current paper is the hyperelastic modelling of a polyurethane adhesive. This material also exhibits a fast decrease of stiffness at low strains which is not well described by widespread hyperleastic models. Without prior knowledge of the work of \citet{yeoh1993}, we developed a different approach to introduce an exponential decrease of a part of the stiffness into the hyperelastic model. In combination with linear viscoelasticity the new hyperelastic model is able to predict the behavior of the investigated adhesive under different loading conditions and strain rates.

\section{Experiments}\label{sec:experiments}

\subsection{Material}
The material "Betaforce 2850L" produced by Dow is a cold-curing, two component polyurethane adhesive. It has a glass transition temperature about -45~$^\circ$C. In specimen manufacture the adhesive was cured one week at room temperature, then exposed to a temperature of 85~$^\circ$C in an oven for one hour, and finally stored at room temperature for at least one week before performing the experiments.

\subsection{Tensile tests}

\begin{figure}[hbtp]
\centering
\includegraphics[height=\picHeight]{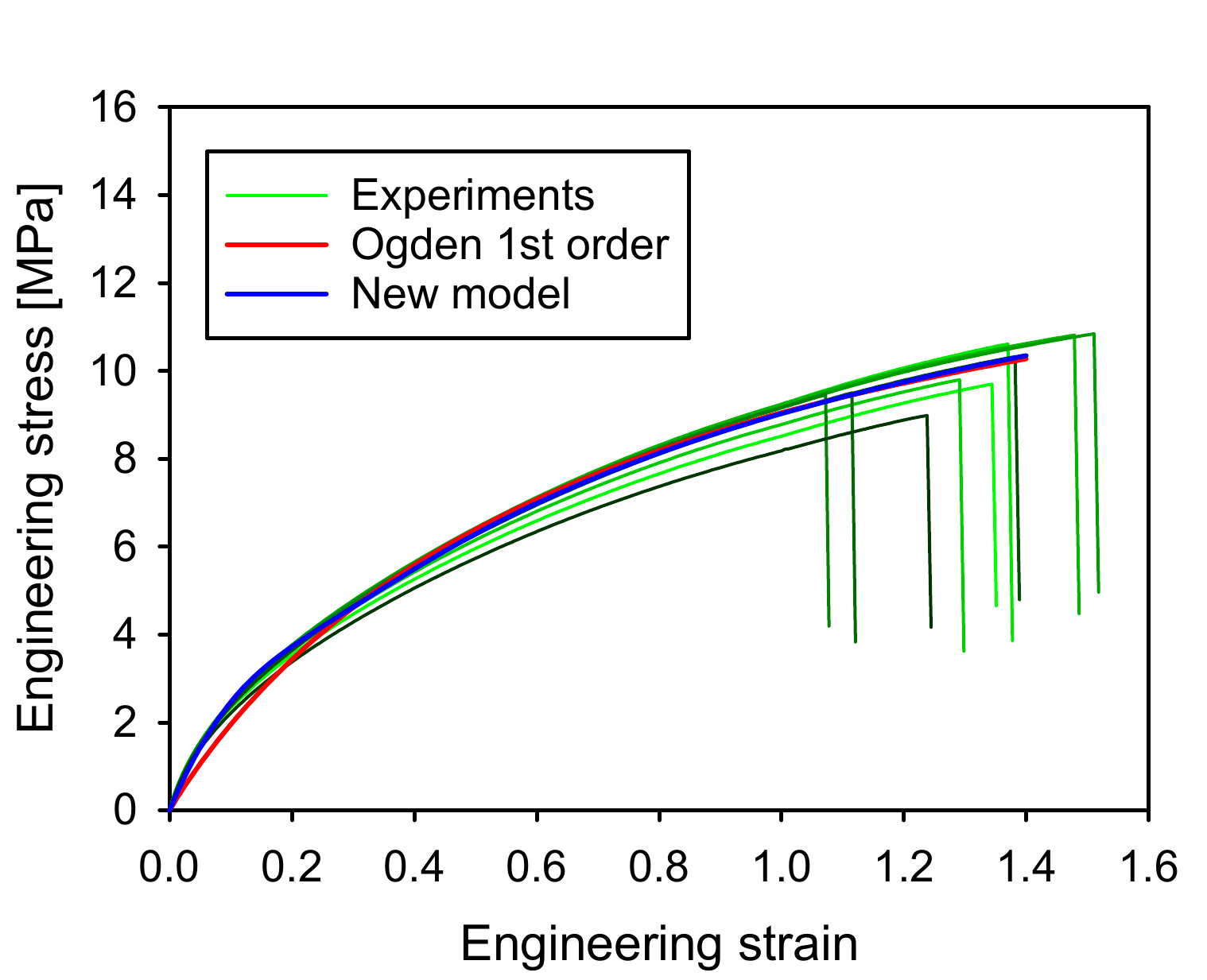}
\caption{Stress-strain curve of uniaxial tensile test at $\dot{\varepsilon}=0.005~$s$^{-1}$ and simulation}
\label{fig:tensile-0p005}
\end{figure}

\begin{figure}[hbtp]
\centering
\includegraphics[height=\picHeight]{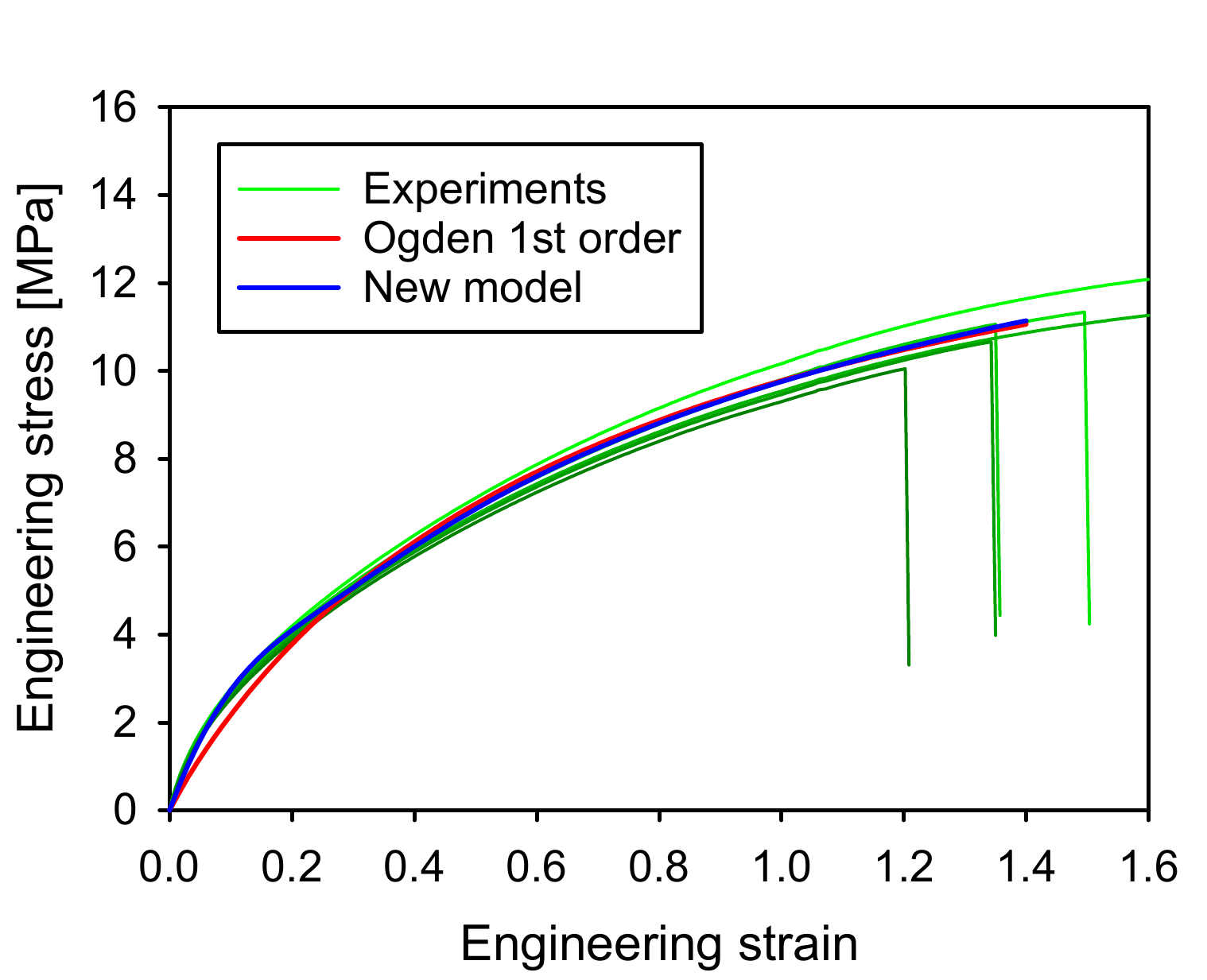}
\caption{Stress-strain curve of uniaxial tensile test at $\dot{\varepsilon}=0.05~$s$^{-1}$ and simulation}
\label{fig:tensile-0p05}
\end{figure}

\begin{figure}[hbtp]
\centering
\includegraphics[height=\picHeight]{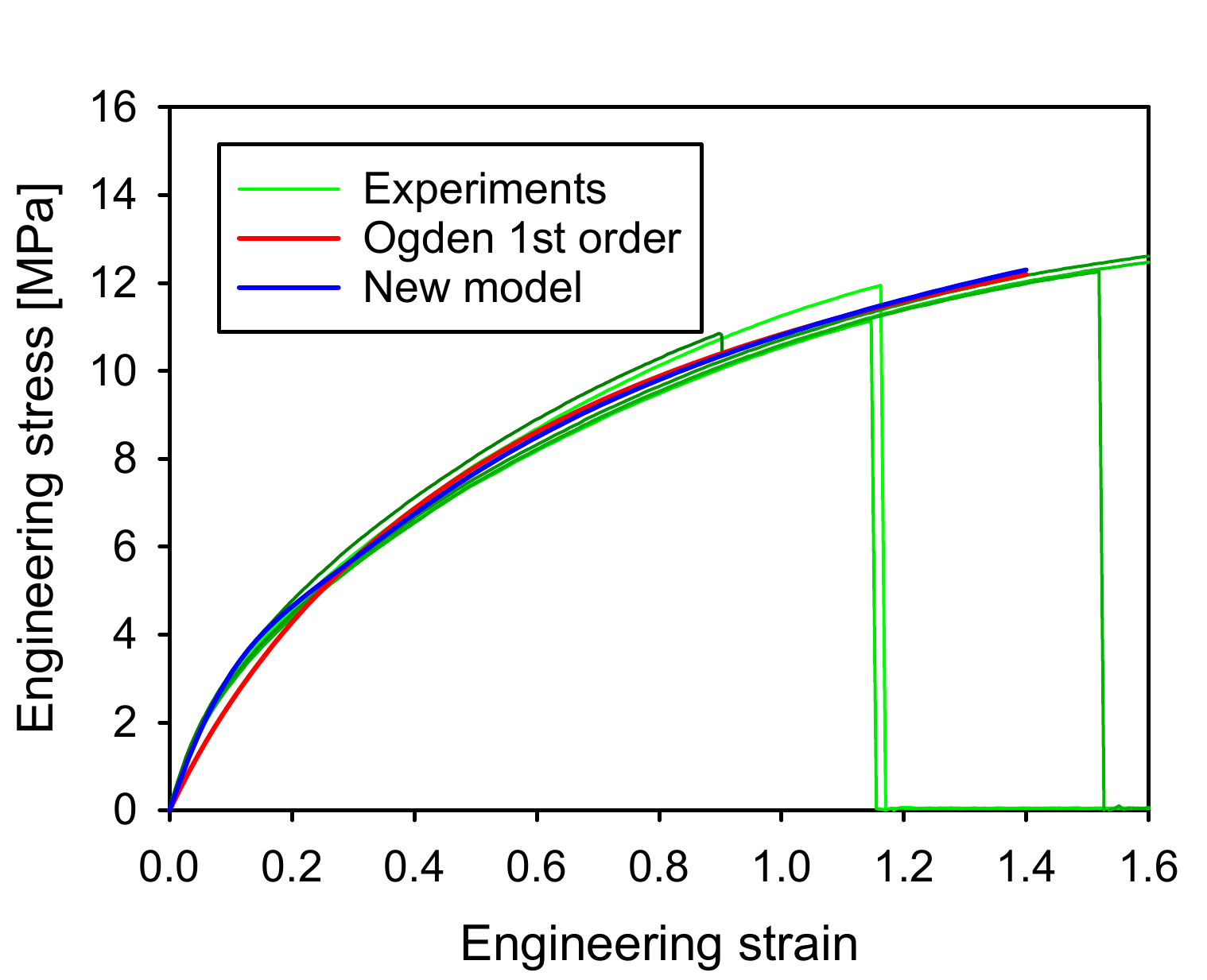}
\caption{Stress-strain curve of uniaxial tensile test at $\dot{\varepsilon}=0.5~$s$^{-1}$ and simulation}
\label{fig:tensile-0p5}
\end{figure}

\begin{figure}[hbtp]
\centering
\includegraphics[height=\picHeight]{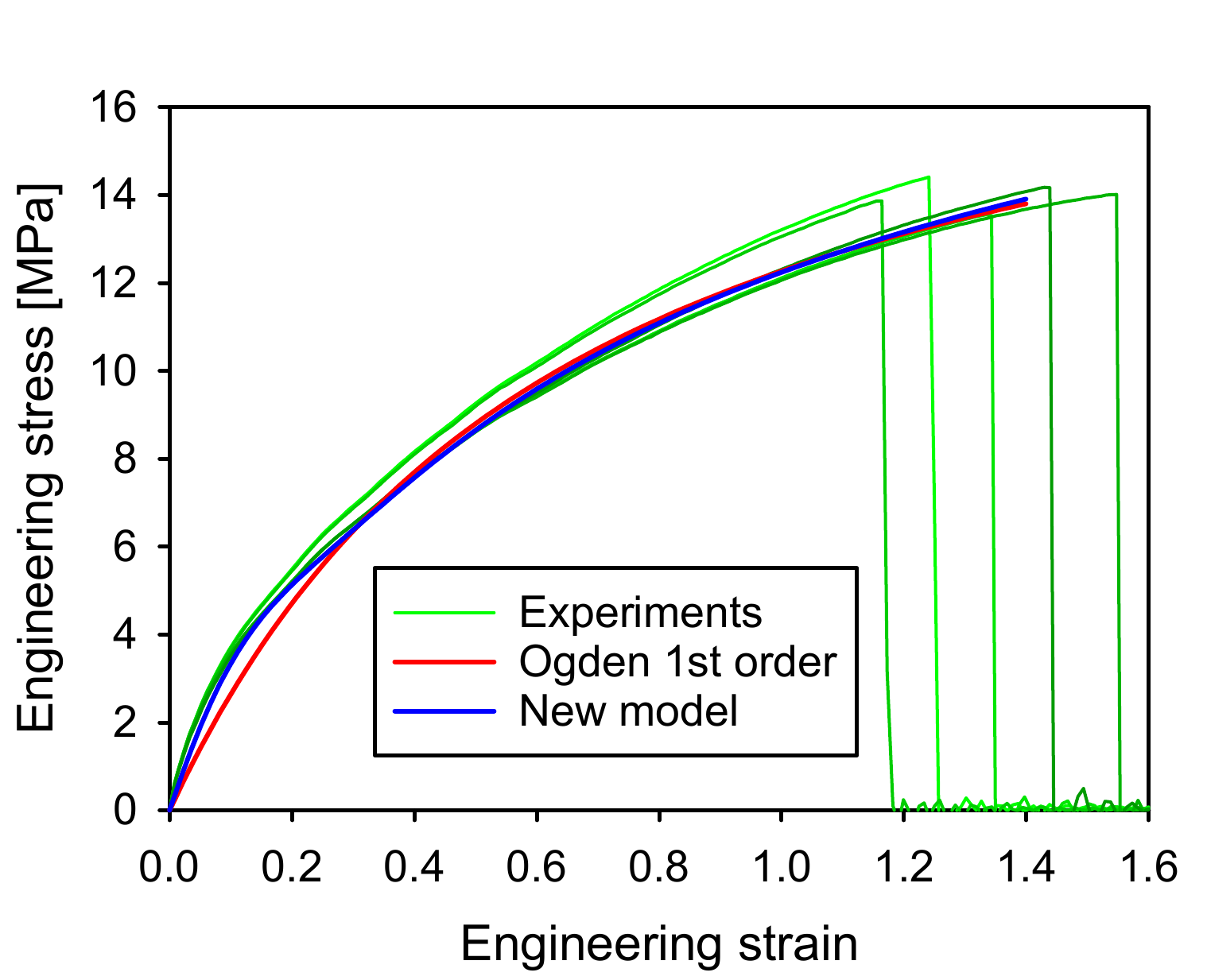}
\caption{Stress-strain curve of uniaxial tensile test at $\dot{\varepsilon}=5~$s$^{-1}$ and simulation}
\label{fig:tensile-5}
\end{figure}

Tensile tests were performed on dog-bone shaped specimens according to DIN EN ISO 527-2 with a parallel specimen length of 80 mm and a cross section of about 10~mm $\times$ 4~mm. The elongation was measured by a mechanical displacement transducer on an initial measurement length of 50~mm. Tests were performed at engineering strain rates of 0.005, 0.05, 0.5, and 5~s$^{-1}$. The resulting stress-strain curves are displayed in figure \ref{fig:tensile-0p005} to \ref{fig:tensile-5}. The figures contain the results of tests on four specimens as well as the stress-strain curves from models which will be discussed in section \ref{sec:val-tensile}.

The stress-strain curves at the four strain rates have a similar shape. At higher strain rates a higher stiffness is visible as it is to be expected for a viscoelastic material. The curves exhibit no S-shape, i.e.~there is no increase of slope before fracture.


\FloatBarrier

\subsection{Compressive tests}

Compressive tests were performed on cylindrical adhesive samples with a diameter about 20~mm and a height about 10~mm. A very thin layer of grease was applied on the top and bottom surface to reduce the effect of friction before the specimen was placed between two pressure plates. A specific setup ensured the parallel alignment of the pressure plates in the servohydraulic test machine. The change of distance between the pressure plates was measured using a mechanical displacement transducer to obtain the strain in the adhesive. Tests were performed at engineering strain rates of 0.0005, 0.005, and 0.05~s$^{-1}$. The specimens were loaded up to an engineering strain of -0.8. No cracks in the specimen were observed during the tests or after unloading.

\begin{figure}[hbtp]
\centering
\includegraphics[height=\picHeight]{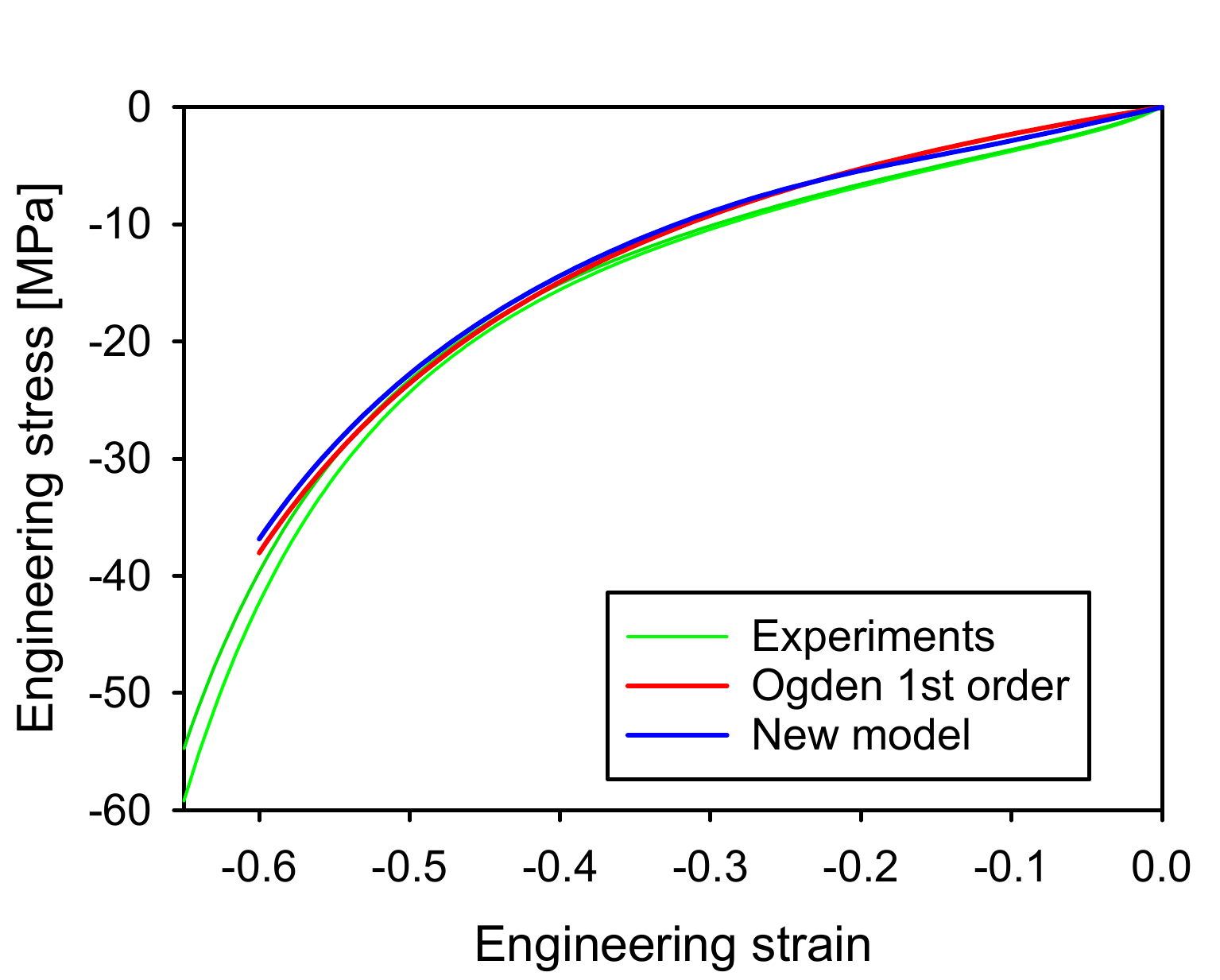}
\caption{Stress-strain curve of uniaxial compression test at $\dot{\varepsilon}=0.0005~$s$^{-1}$ and simulation}
\label{fig:compression-0p0005}
\end{figure}
\begin{figure}[hbtp]
\centering
\includegraphics[height=\picHeight]{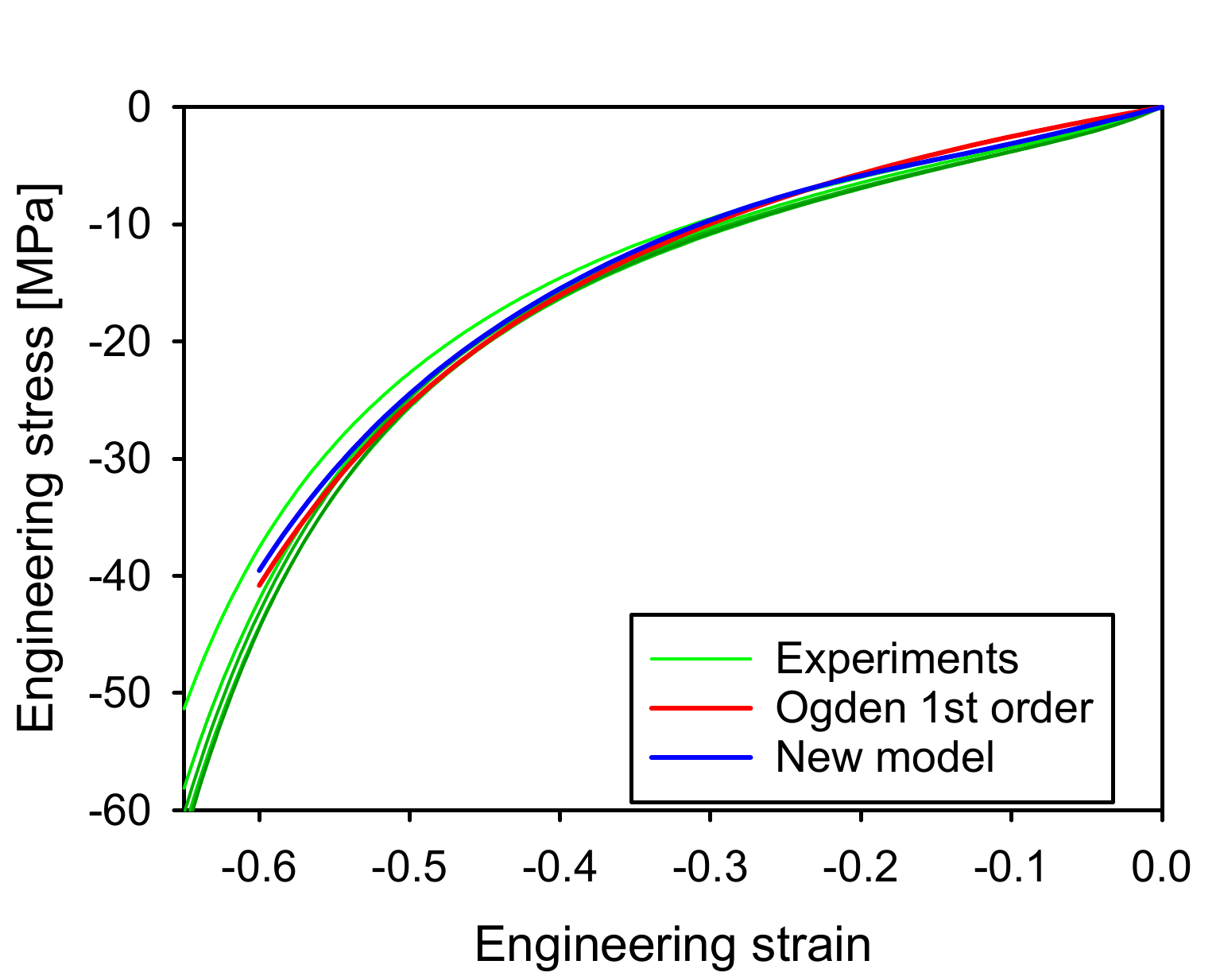}
\caption{Stress-strain curve of uniaxial compression test at $\dot{\varepsilon}=0.005~$s$^{-1}$ and simulation}
\label{fig:compression-0p005}
\end{figure}
\begin{figure}[hbtp]
\centering
\includegraphics[height=\picHeight]{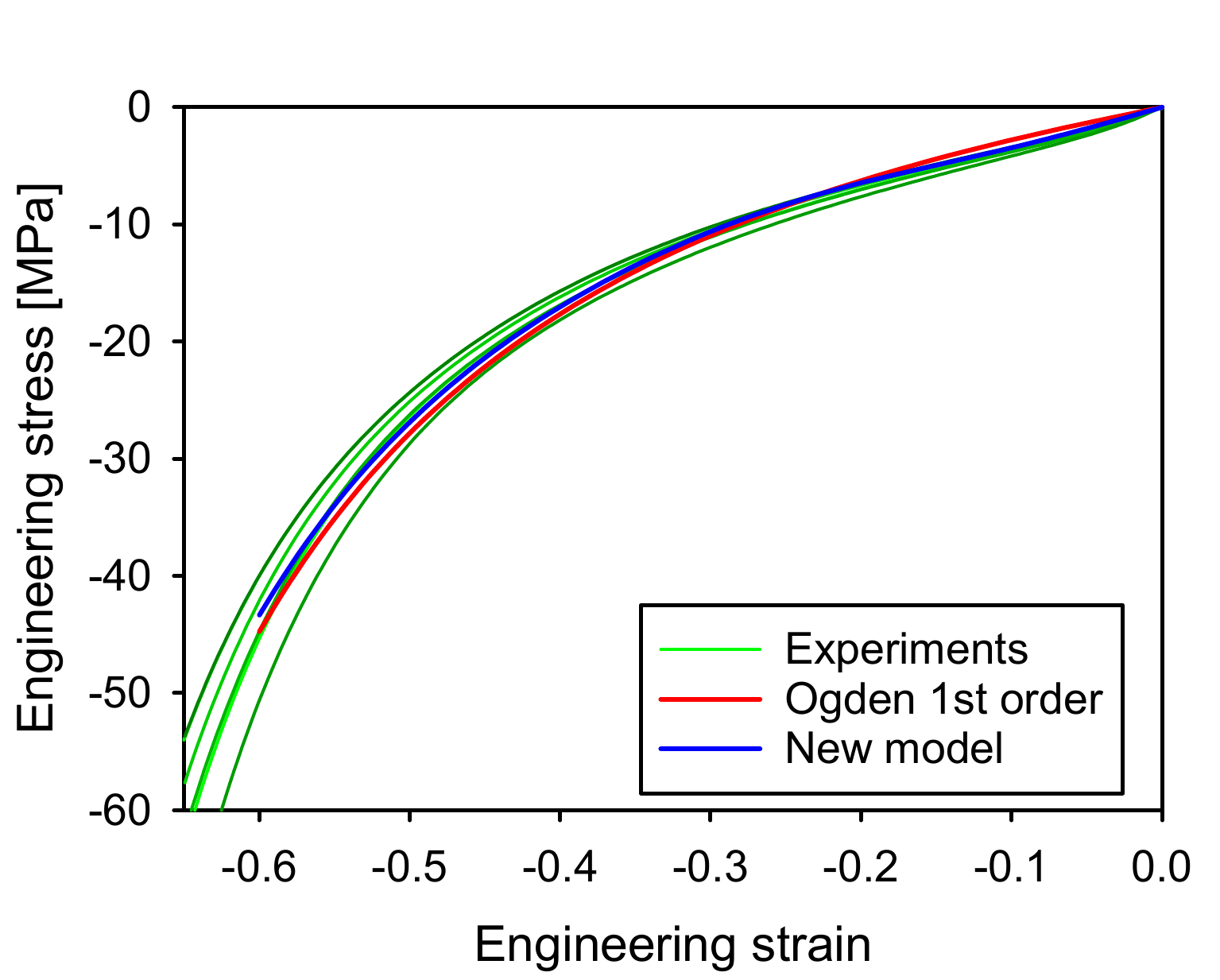}
\caption{Stress-strain curve of uniaxial compression test at $\dot{\varepsilon}=0.05~$s$^{-1}$ and simulation}
\label{fig:compression-0p05}
\end{figure}

\begin{figure}[hbtp]
\centering
\includegraphics[width=0.49\textwidth]{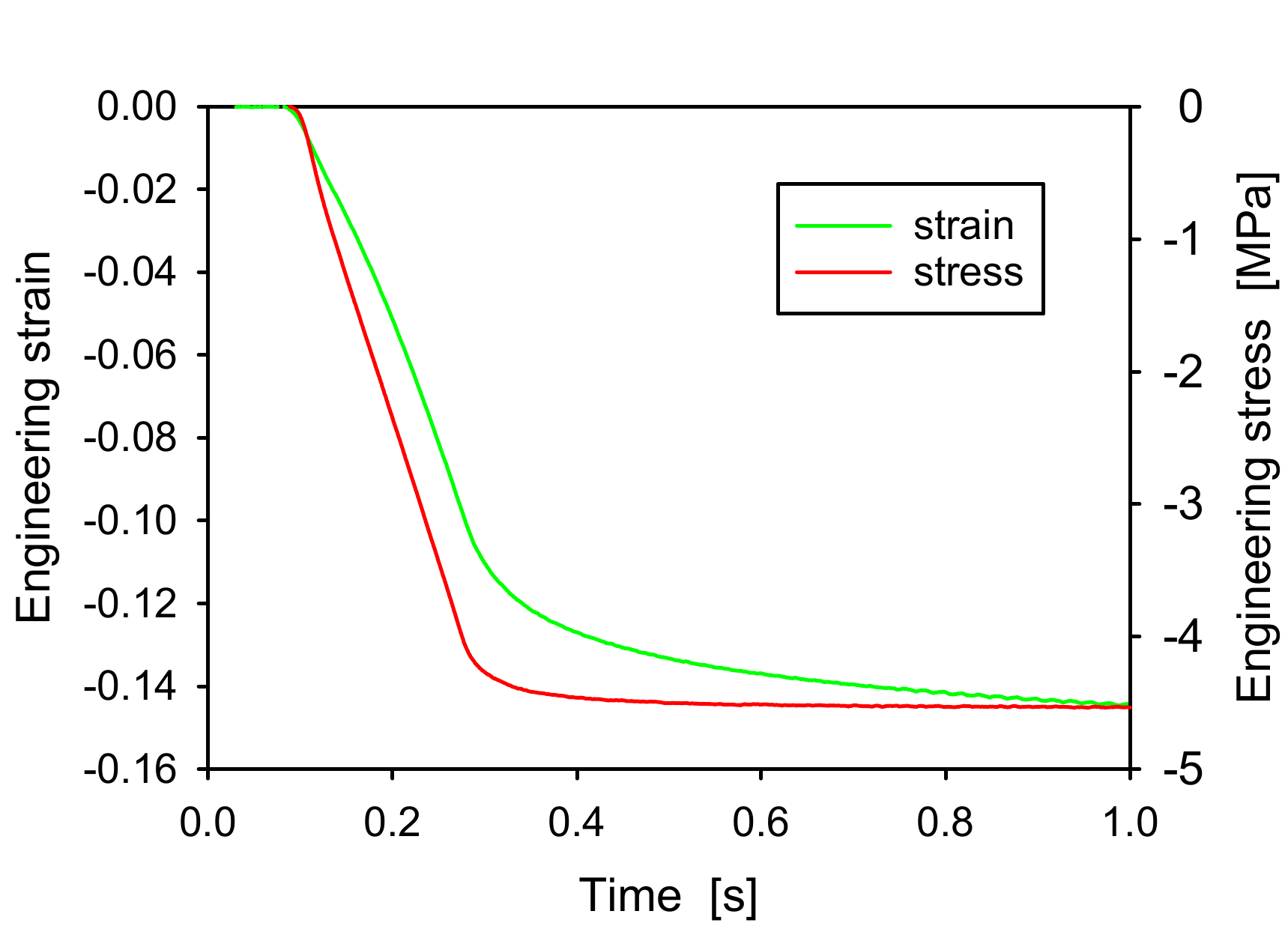}
\includegraphics[width=0.49\textwidth]{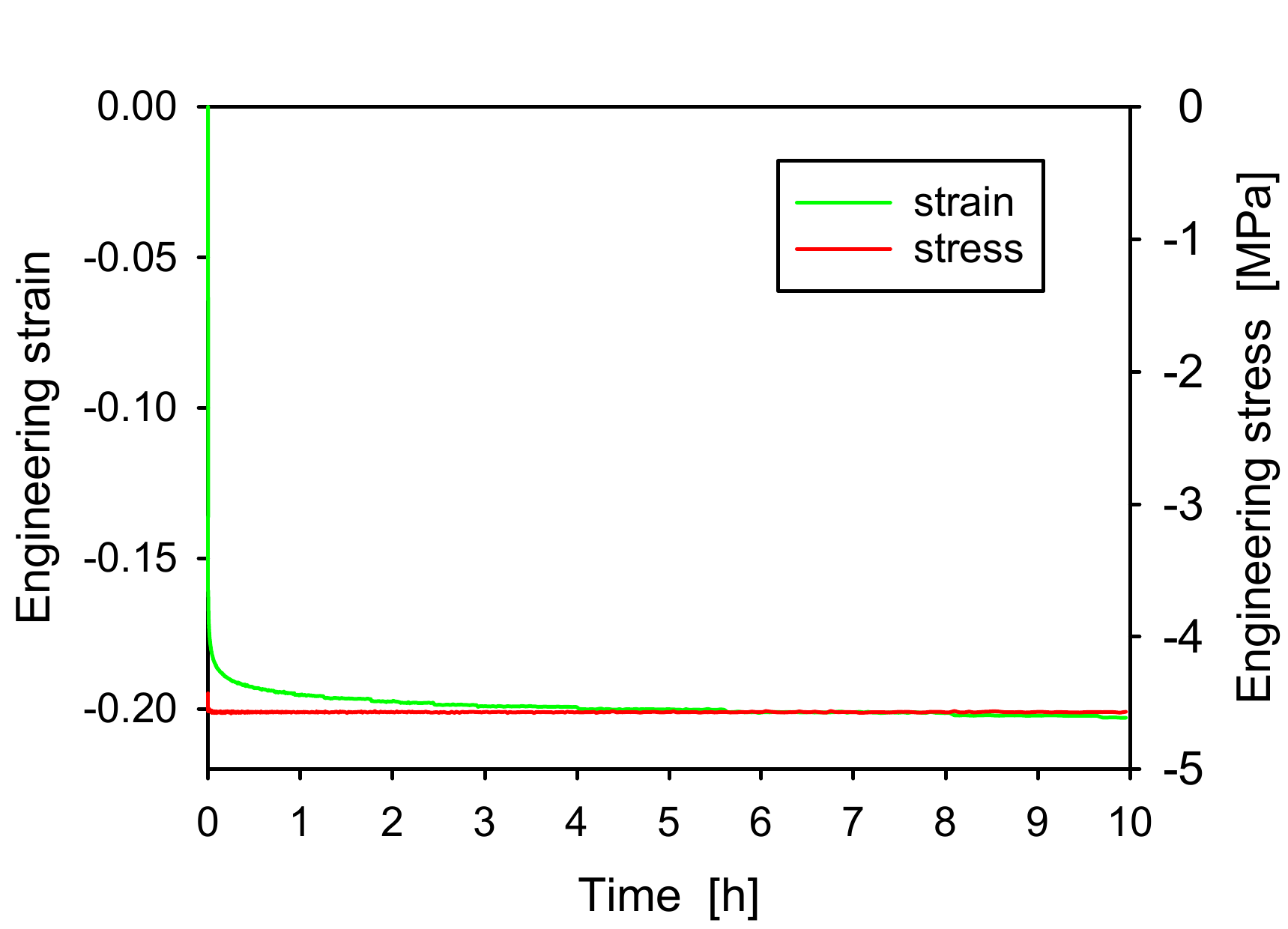}
\caption{Stress and strain-time curves of uniaxial compression creep test at 4.6~MPa stress level, left: start of test, right: entire test}
\label{fig:creep46}
\end{figure}

\begin{figure}[hbtp]
\centering
\includegraphics[width=0.49\textwidth]{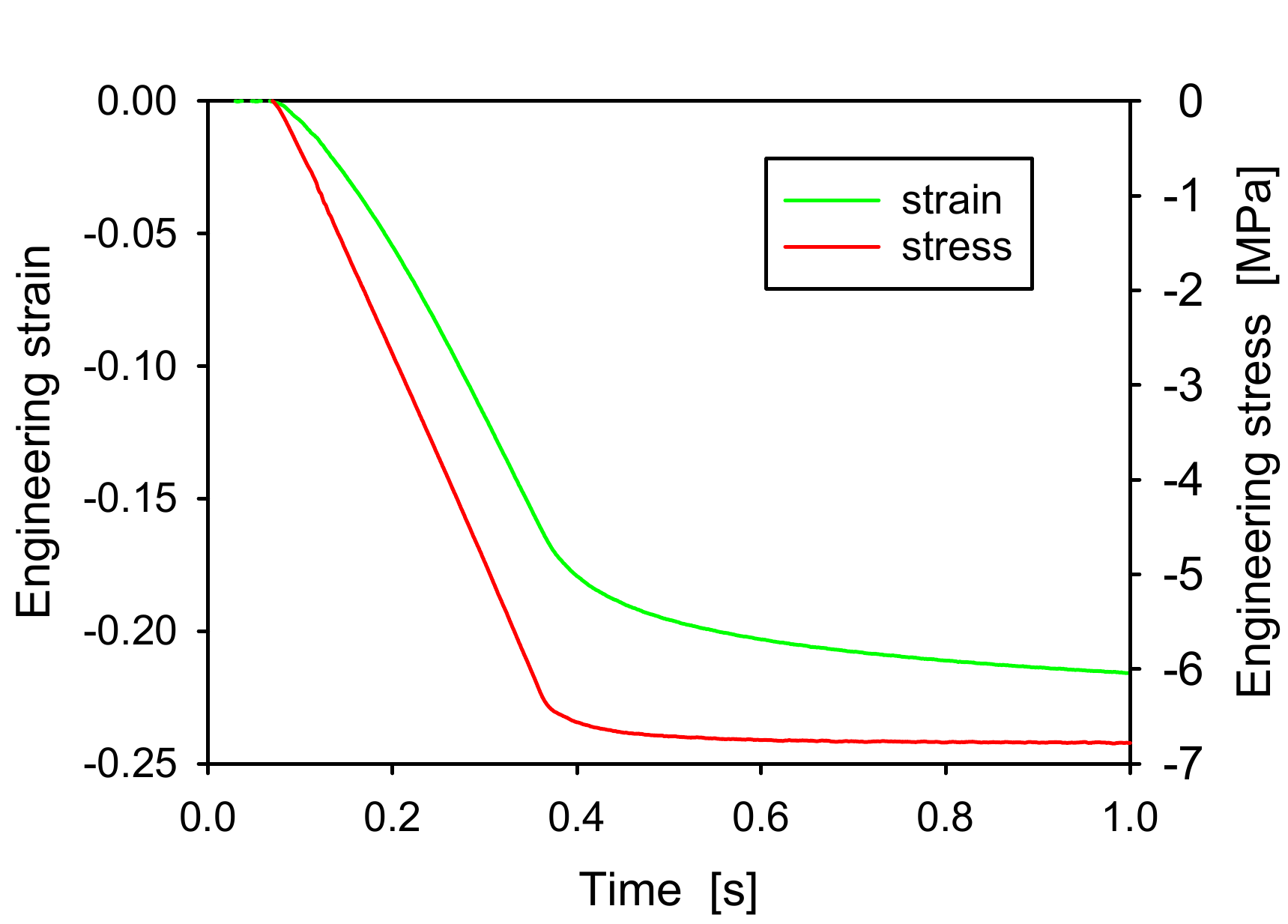}
\includegraphics[width=0.49\textwidth]{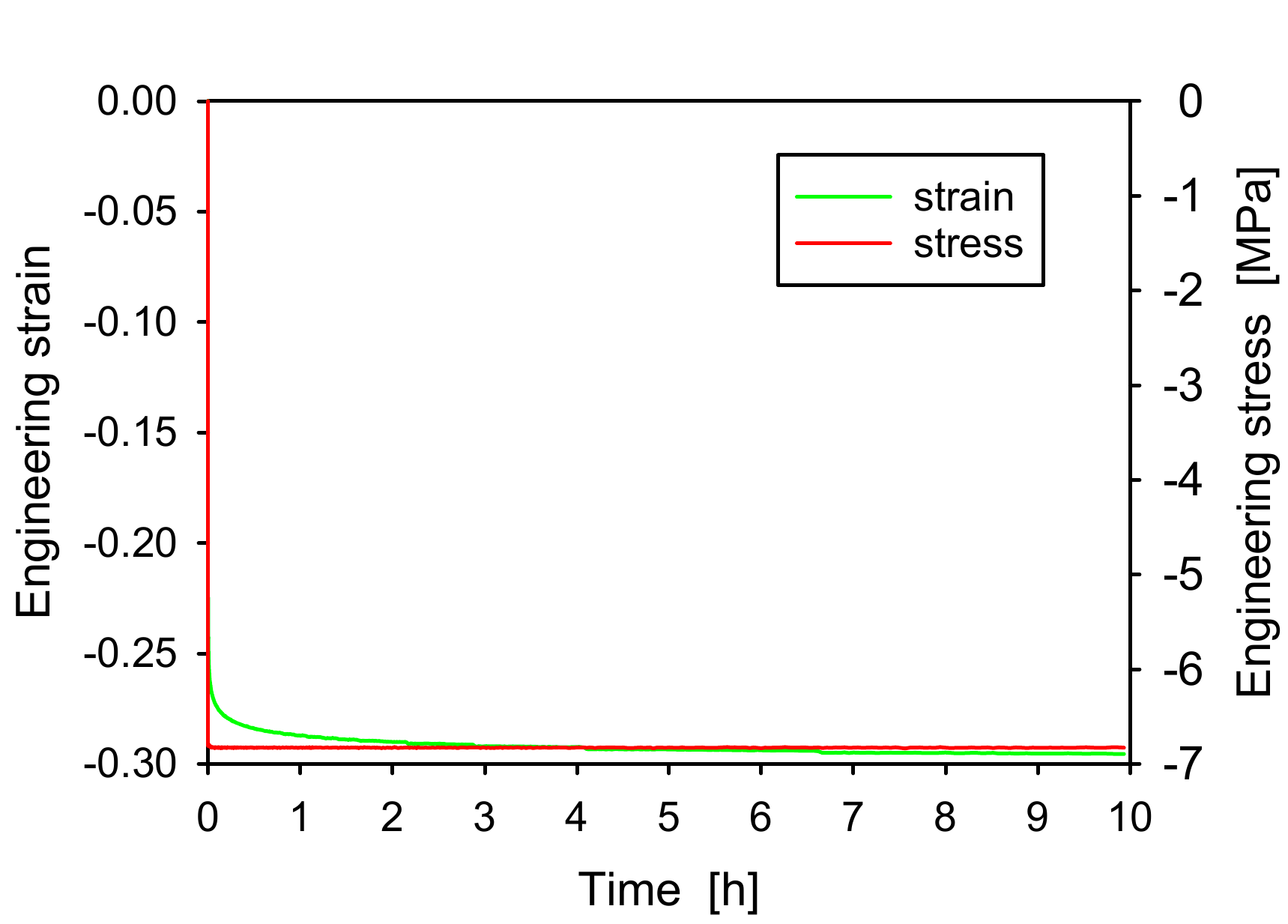}
\caption{Stress and strain-time curves of uniaxial compression creep test at 6.9~MPa stress level, left: start of test, right: entire test}
\label{fig:creep69}
\end{figure}

Additionally, compression tests were performed as creep tests at two load levels (4.6 and 6.9~MPa, figures \ref{fig:creep46} and \ref{fig:creep69}). The force was applied within 0.2~s and then kept constant over 10~h duration.


\FloatBarrier

\subsection{Lap shear tests}

\begin{figure}[hbtp]
\centering
\includegraphics[height=50mm]{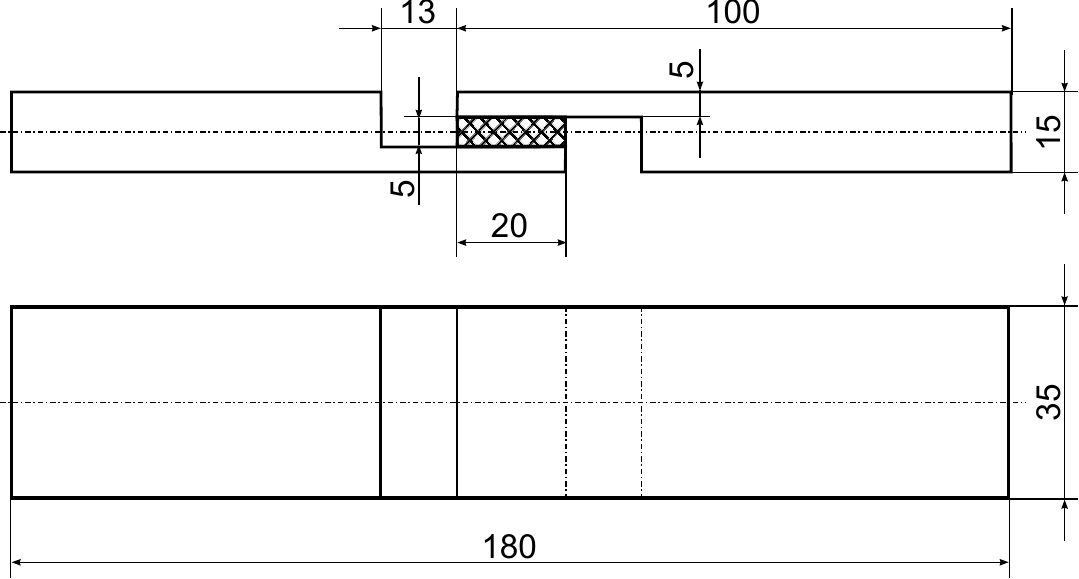}
\caption{Geometry of lap shear specimen according to DIN 6701-3, dimensions in mm: side view (top) and top view (bottom)}
\label{fig:lap-shear}
\end{figure}

Thick adherend lap shear tests were performed according to DIN 6701-3. The specimen geometry is sketched in figure \ref{fig:lap-shear}. The adherends were made from aluminium AW~6082 and pretreated by wiping using isopropanol. The specimens were clamped at a free length of 80~mm and loaded at a constant crosshead displacement velocity according to nominal strain rates of 0.01, 1, and 5~s$^{-1}$. A mechanical displacement transducer was used to obtain a local measurement of elongation on a measurement length of 35~mm. Shear stress - shear strain curves were evaluated by considering the nominal shear strain based on the local displacement measurement and the nominal shear stress. This evaluation neglects the stress inhomogeneity in the adhesive layer and the bending of the thick adherends.

\begin{figure}[hbtp]
\centering
\includegraphics[height=\picHeight]{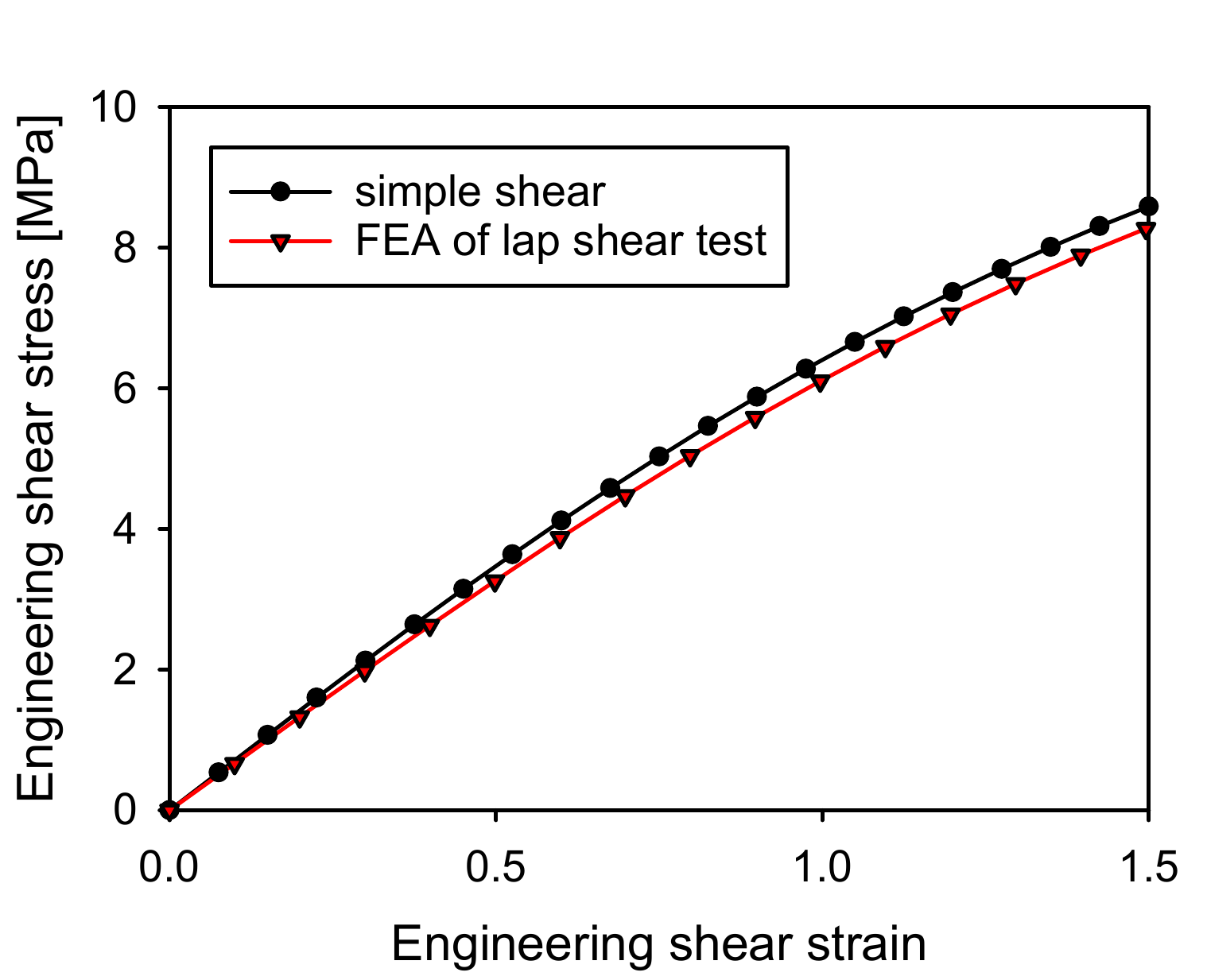}
\caption{Comparison between evaluation of simulated test and material model behaviour in simple shear}
\label{fig:lap-shear-idealization}
\end{figure}

It should be noted that neglecting effects of inhomogeneity and bending in lap shear tests is usually inappropriate in case of stiff adhesive layers. To check whether this simplification causes a significant error, a finite element analysis of the lap shear test was performed using the finite element software Abaqus \cite[]{abaqus}. The specimen was modelled in three dimensions with linear, reduced integration, quadrilateral elements and an element edge length ranging between 0.5 and 1~mm. Hybrid elements were used in the adhesive layer employing a first order Ogden model with material parameters given in table \ref{tab:parameters}. Not the entire 180~mm length of the specimen was modelled, but only the 80~mm between the clamps in the test. Displacement boundary conditions were prescribed at the model faces at the two clamps. The displacements were evaluated at the positions where the displacement transducer is fastened in the experiment. Engineering stress and strain were calculated from these displacements and from the reaction forces in the simulation in the same way as the measurements in the experiment were evaluated. Figure \ref{fig:lap-shear-idealization} shows the result as a red curve. For comparison, the black curve displays the behaviour of the same material model under ideal simple shear conditions which was obtained by performing a one-element-test. The difference is quite small, at a strain of 1.5 it is less than 4~\% of the stress.

\begin{figure}[hbtp]
\centering
\includegraphics[height=\picHeight]{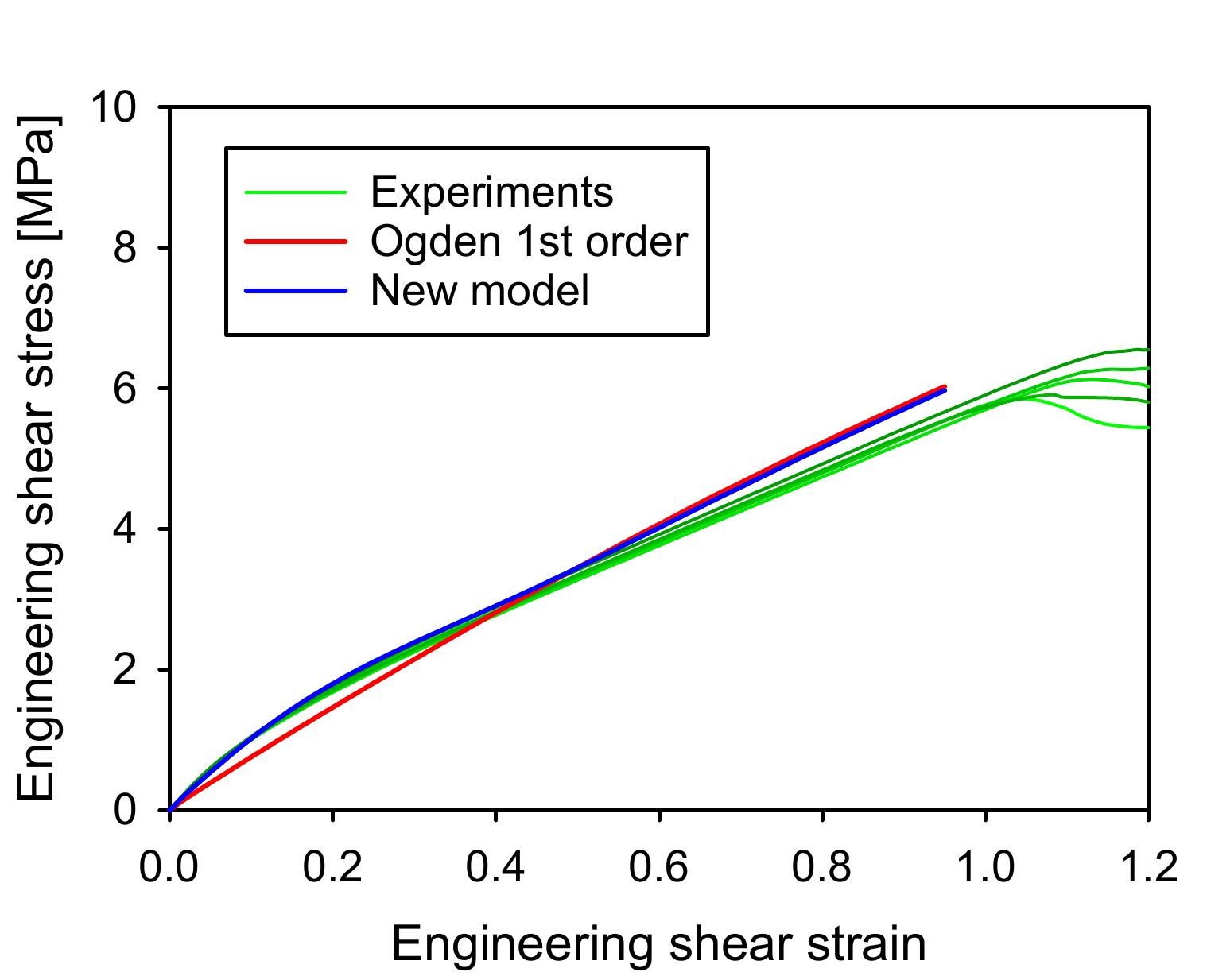}
\caption{Shear stress-strain curve of lap shear test at $\dot{\varepsilon}=0.01~$s$^{-1}$ and simulation}
\label{fig:shear-0p01}
\end{figure}

\begin{figure}[hbtp]
\centering
\includegraphics[height=\picHeight]{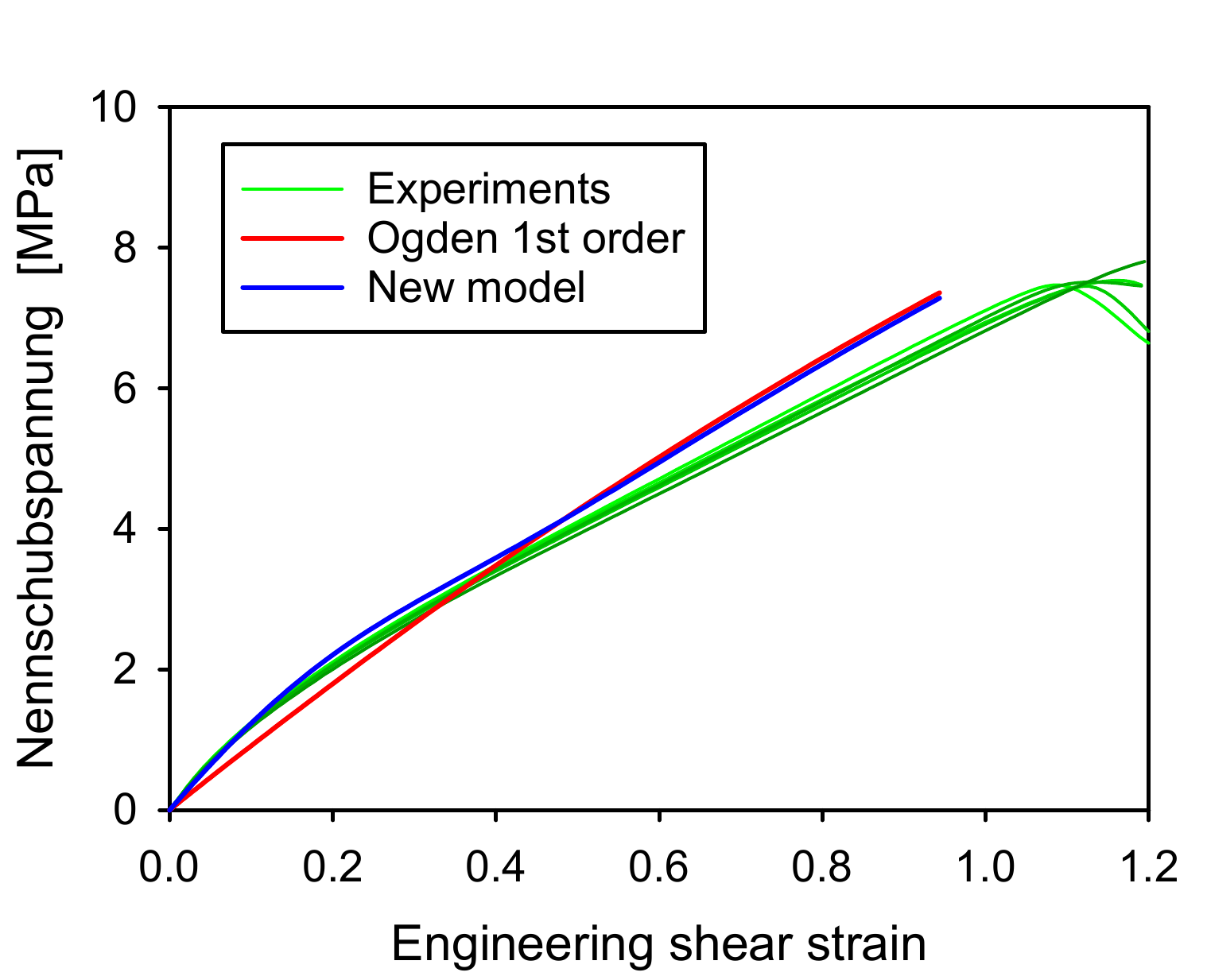}
\caption{Shear stress-strain curve of lap shear test at $\dot{\varepsilon}=1~$s$^{-1}$ and simulation}
\label{fig:shear-1}
\end{figure}

\begin{figure}[hbtp]
\centering
\includegraphics[height=\picHeight]{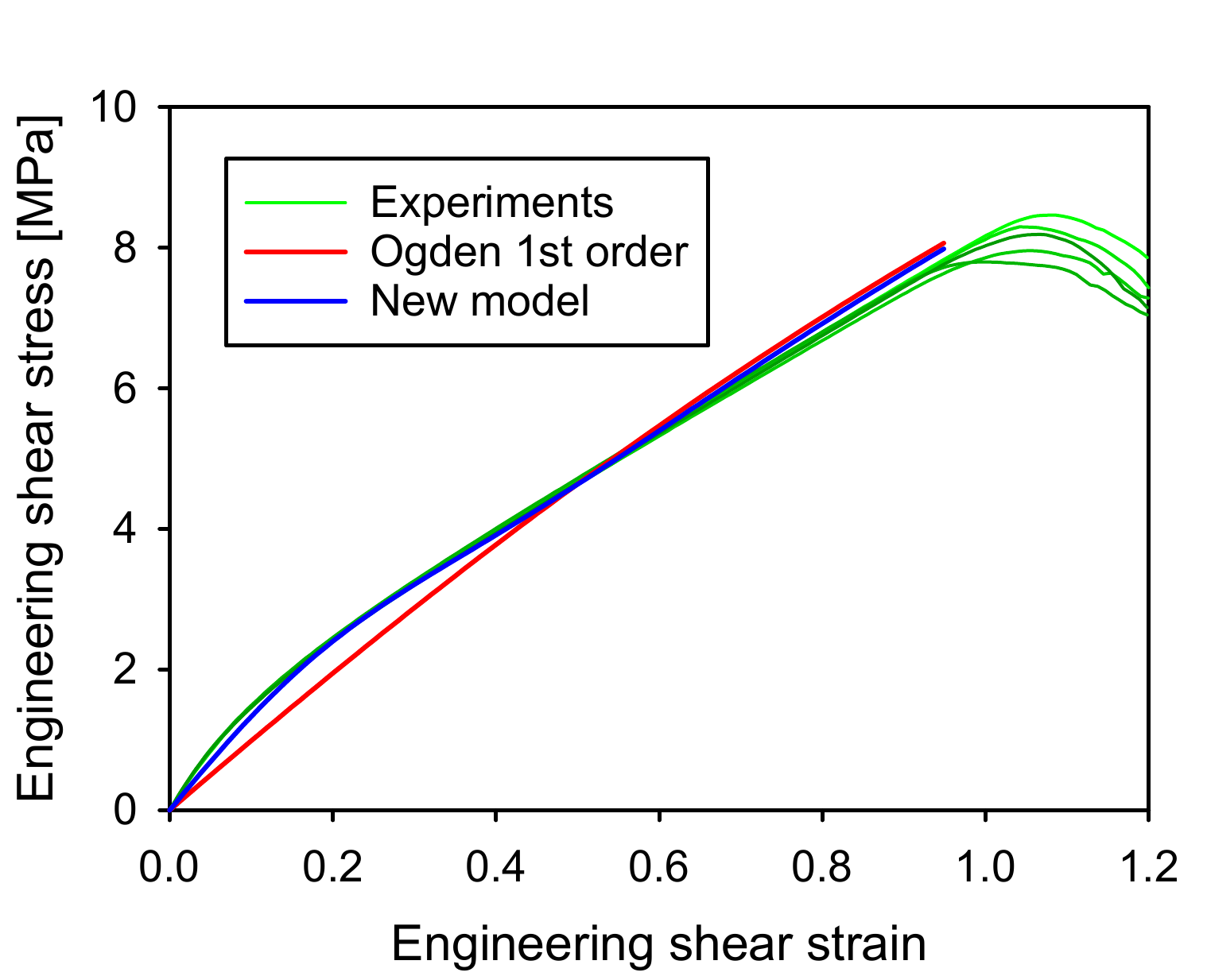}
\caption{Shear stress-strain curve of lap shear test at $\dot{\varepsilon}=5~$s$^{-1}$ and simulation}
\label{fig:shear-5}
\end{figure}

The results of the lap shear tests at three different strain rates are displayed in figure \ref{fig:shear-0p01} to \ref{fig:shear-5}. The increase in stiffness at increasing strain rate due to viscoelasticity can clearly be seen in comparison of figure \ref{fig:shear-0p01} and \ref{fig:shear-1} which differ by a factor 100 in strain rate. Concerning the shape of the stress-strain curves, at first they show a decrease of slope until a strain of about 0.2 is reached. Afterwards, the curve is approximately linear, until the stress comes close to its maximum, when probably damage processes cause a decrease of slope. The curves of the simulations are topic of section \ref{sec:val-shear}.


\FloatBarrier

\subsection{Butt joint tests}\label{sec:butt}

Two kinds of butt joint specimens were manufactured by bonding two cylindrical steel adherends with 15~mm diameter by an adhesive layer of either 2 or 5~mm thickness. A tensile load was applied at an engineering strain rate of 0.005~s$^{-1}$. The relative displacement of the two steel adherends was measured using a displacement transducer to obtain the strain of the adhesive layer. Due to the relative compliance of the adherends and the adhesive layer, the contribution of the steel deformation to this measurement is less than 0.1~\% of its value. The manufacture and testing of the butt joint specimens was performed at the Laboratorium f{\"u}r Werkstoff- und F{\"u}getechnik of Paderborn University.

Since the elastic modulus of the adherends is higher than the modulus of the adhesive by more than a factor $10^4$, the adherends show only very little lateral contraction and restrict the lateral contraction of the adhesive. This restriction causes the adhesive in the butt joint to exhibit a stiffer behaviour than the adhesive in a uniaxial tensile test. Since lateral contraction of the adhesive in the butt joint can mainly happen close to the free surface of the adhesive, the amount of restriction and thus the stiffness depends on the ratio of adhesive layer thickness to specimen diameter. With decreasing aspect ratio of adhesive layer thickness to specimen diameter, the test approaches the limit of the restricted tension test and the stiffness increases \citep{lindley1979}.

Figure \ref{fig:shear-5} displays the engineering stress-strain curves obtained from the tests. The curves of the specimens with thinner adhesive layer show a higher slope than the curves of specimens with thicker layer due to the higher amount of restriction of lateral contraction.

\begin{figure}[hbtp]
\centering
\includegraphics[height=\picHeight]{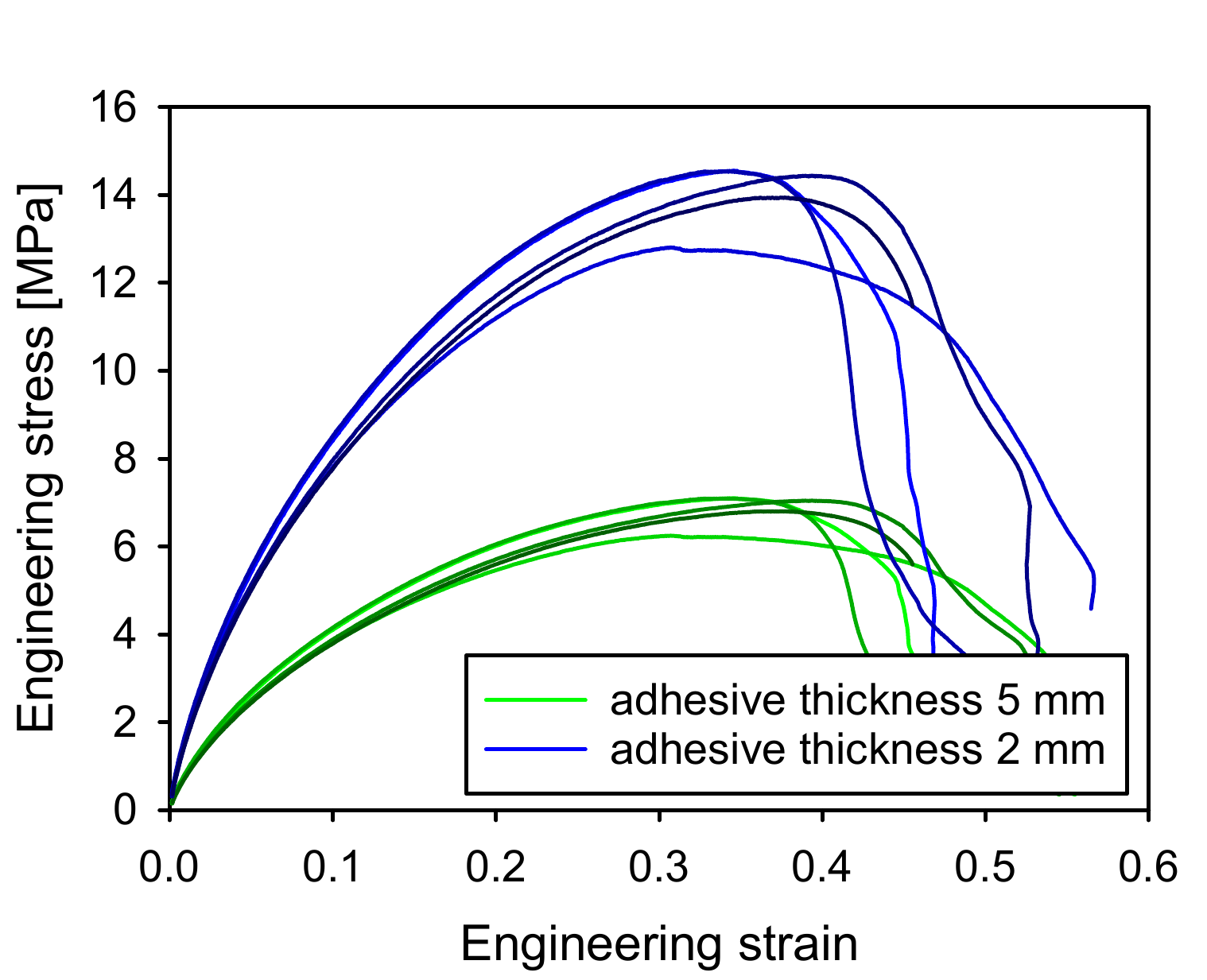}
\caption{Engineering stress-strain curve of but joint tests with 2 and 5~mm adhesive layer thickness}
\label{fig:butt-totl}
\end{figure}

\section{Application of established hyperelastic models}\label{sec:established}

In this section we will check the accuracy of a fit of the experimental data from the tensile test at one arbitrarily chosen strain rate by some well established hyperelastic models which are available in several commercial finite element programs. Because the material is nearly incompressible and no loadings with a high ratio of hydrostatic stress to the von Mises equivalent stress are considered, only incompressible models are tested in this section. The first subsection shows models which are unconditionally stable, i.e.~fulfill the Drucker stability criterion for any strain state. An example of these kind of models are polynomial models of first order. Higher order models often allow a more accurate description of a single test, but care must be taken since their range of stability might be limited and their extrapolation to different kinds of loading may be questionable.

\subsection{Unconditionally stable models}

Table \ref{tab:stable} lists the strain energy density functions of the unconditionally stable models which have been used to fit the results of the tensile test at a strain rate of 0.005~s$^{-1}$. They contain the reduced polynomial, full polynomial and Ogden model of first order. The Arruda-Boyce model contains terms of higher order, but in a special form with only two free parameters, so that it ensures stability. $\bar{I}_1,\bar{I}_2$ denote the deviatoric strain invariants, which depend on the principal stretches $\lambda_1,\lambda_2,\lambda_3$ as follows:
\begin{gather}
J = \lambda_1\,\lambda_2\,\lambda_3, \qquad \bar{\lambda}_i = J^{-1/3}\,\lambda_i \nonumber\\
\bar{I}_1 = \bar{\lambda}_1^2 + \bar{\lambda}_2^2 + \bar{\lambda}_3^2, \qquad \bar{I}_2 = \bar{\lambda}_1^{-2} + \bar{\lambda}_2^{-2} + \bar{\lambda}_3^{-2} \label{eq:invariants}
\end{gather}

\begin{table}[htbp]
\centering
\caption{Tested stable hyperelastic models \label{tab:stable}}
\begin{tabular}{cc}
\hline \rule[-2ex]{0pt}{5.5ex} Model & Potential \\ 
\hline \rule[-2ex]{0pt}{5.5ex} neo-Hooke & $C_{10}\,(\bar{I}_1-3)$ \\ 
\hline \rule[-2ex]{0pt}{5.5ex} Mooney-Rivlin & $C_{10}\,(\bar{I}_1-3) + C_{01}\,(\bar{I}_2-3)$ \\ 
\hline \rule[-2ex]{0pt}{5.5ex} Ogden 1st order & $\frac{2\,\mu}{\alpha^2} \left(\bar{\lambda}_1^\alpha + \bar{\lambda}_2^\alpha + \bar{\lambda}_3^\alpha - 3\right)$ \\ 
\hline \rule[-2ex]{0pt}{5.5ex} Arruda-Boyce & $\mu \,\left[ \frac{1}{2} (\bar{I}_1-3) + \frac{1}{20\lambda_m^2} (\bar{I}_1^2-9) + 
								\frac{11}{1050\lambda_m^4} (\bar{I}_1^3-27) \right.$ \\
								 & $+ \left. \frac{19}{7000\lambda_m^6} (\bar{I}_1^4-81) + \frac{519}{673750\lambda_m^8} (\bar{I}_1^5-243) \right]$ \\ 
\hline 
\end{tabular} 
\end{table}

\begin{figure}[hbtp]
\centering
\includegraphics[height=\picHeight]{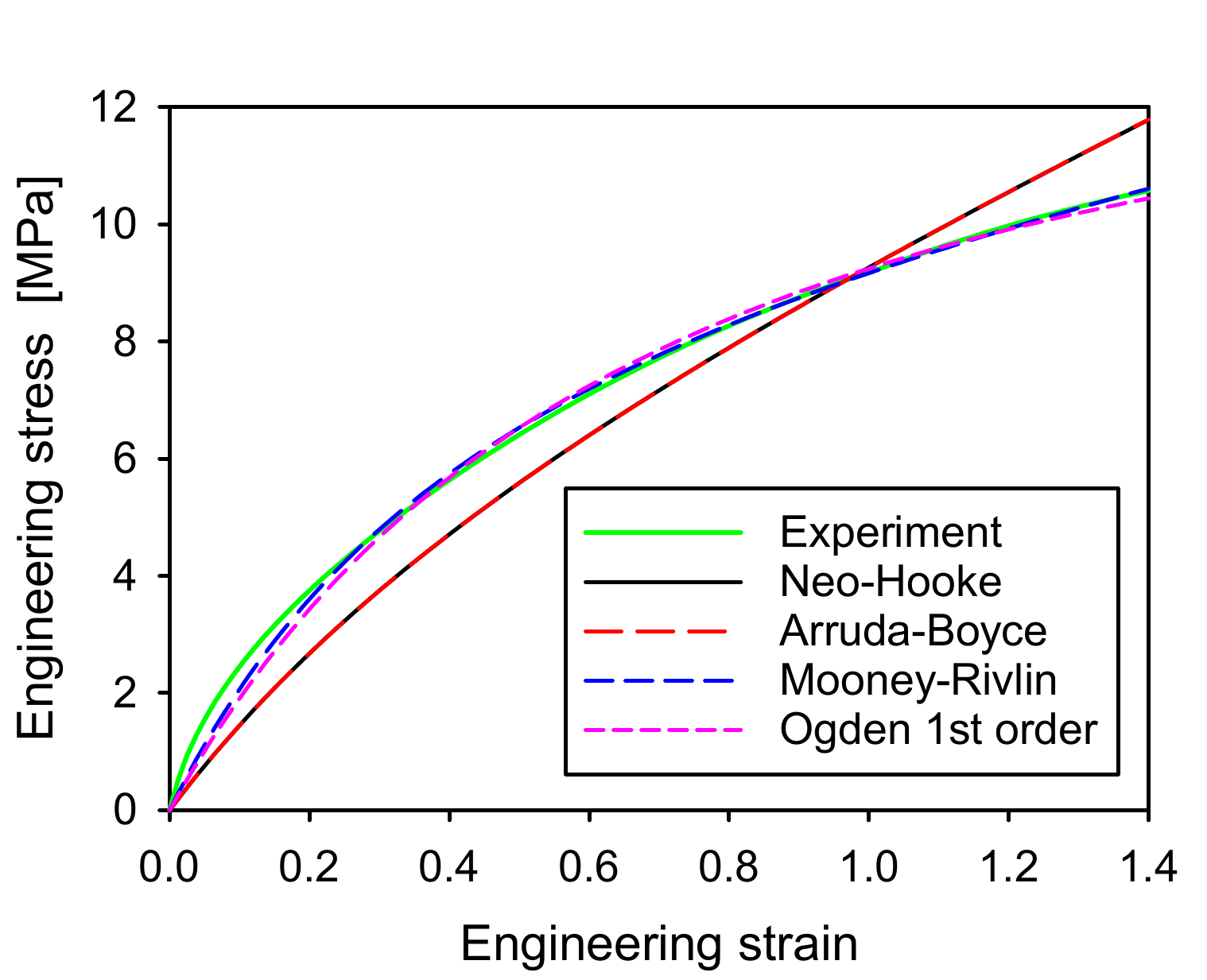}
\caption{Fit of tensile test data by four stable hyperelastic models}
\label{fig:stable}
\centering
\includegraphics[height=\picHeight]{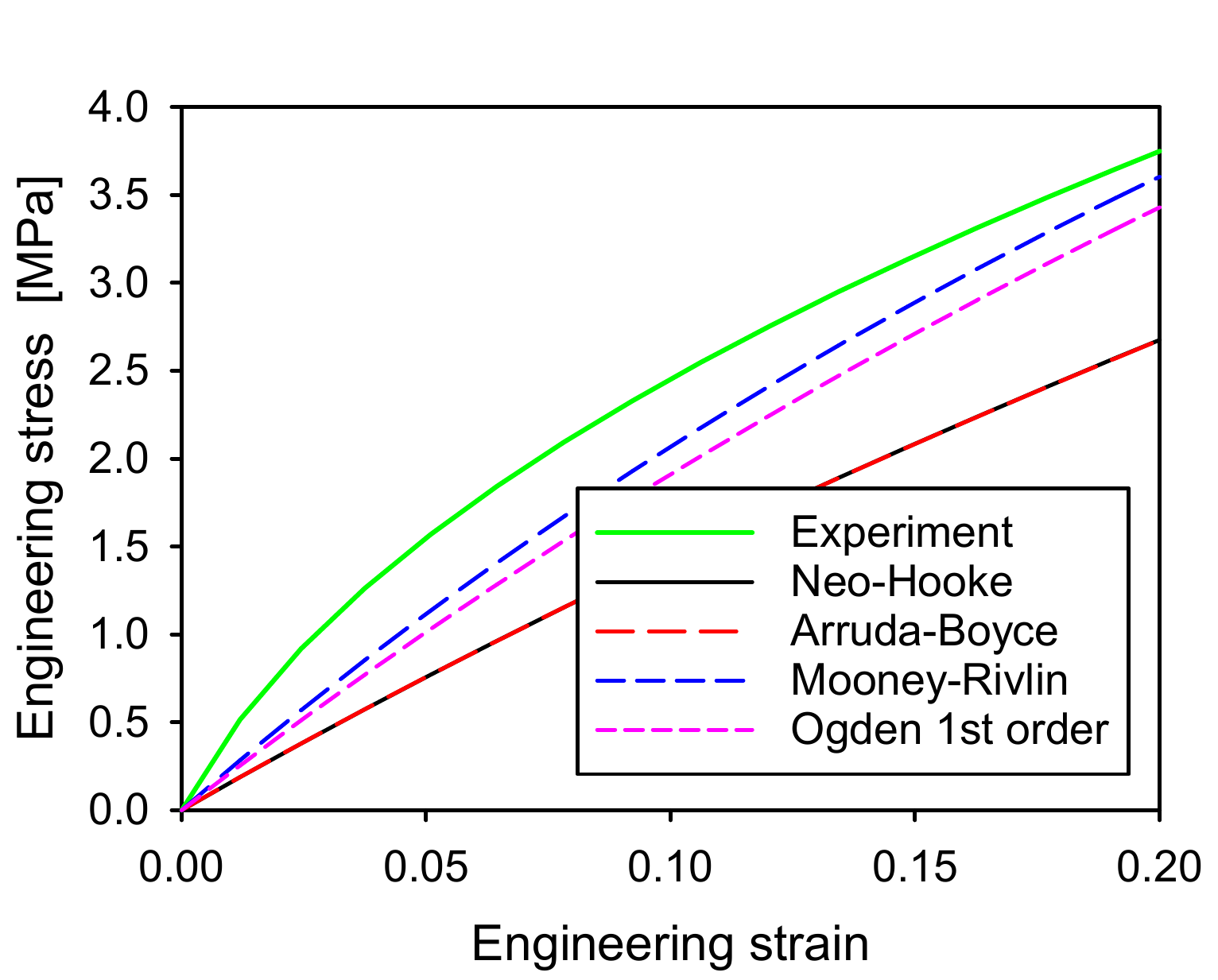}
\caption{Fit of tensile test data by four stable hyperelastic models, result at low strains}
\label{fig:stable-start}
\end{figure}

The different models have been fitted to one typical stress-strain curve. The results of the parameter optimizations are displayed in figure \ref{fig:stable}. The Mooney-Rivlin model and the first order Ogden model describe the results of the tensile tests quite well. However, a closer look at the model behavior at moderately low strains (figure \ref{fig:stable-start}) shows that the stiffness of the material is initially about 50~\% higher than the stiffness of the models.

\FloatBarrier

\subsection{Conditionally stable models}

\begin{figure}[hbtp]
\centering
\includegraphics[height=\picHeight]{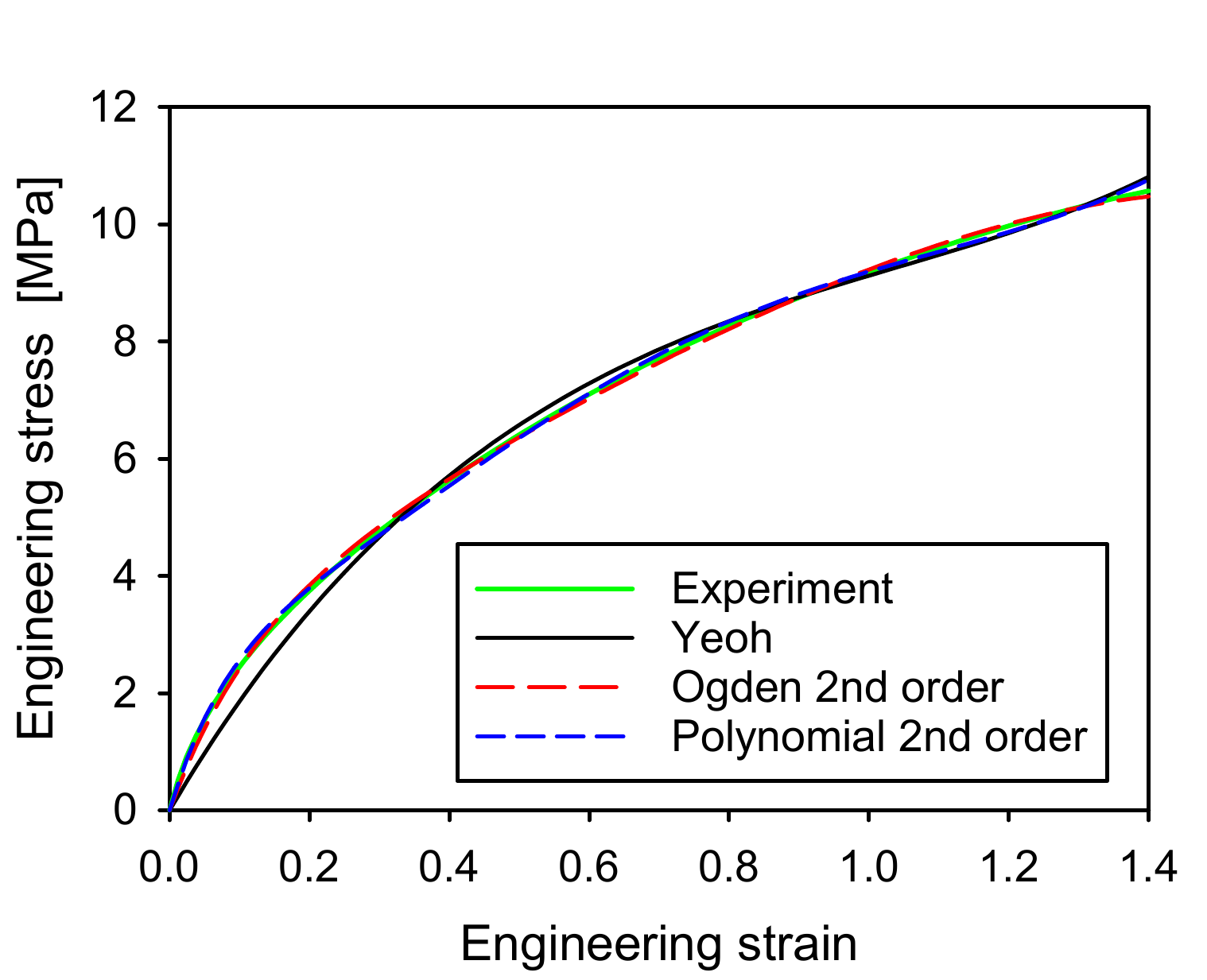}
\caption{Fit of tensile test data by three conditionally stable hyperelastic models}
\label{fig:unstable}
\end{figure}

The deviation between experiment and model for small strains can be decreased by choosing a model of higher order. Figure \ref{fig:unstable} shows the results of fits using the models listed in table \ref{tab:unstable}. While the Yeoh model (without additional exponential term) still shows a significant deviation from the test results at low strains, both polynomial and Ogden model of second order provide a good fit over the entire strain range.

\begin{table}[htbp]
\centering
\caption{Tested conditionally stable hyperelastic models \label{tab:unstable}}
\begin{tabular}{cc}
\hline \rule[-2ex]{0pt}{5.5ex} Model & Potential \\ 
\hline \rule[-2ex]{0pt}{5.5ex} polynomial 2nd order & $\sum\limits_{i+j=1}^{2} C_{ij}\,(\bar{I}_1-3)^i\,(\bar{I}_2-3)^j$ \\ 
\hline \rule[-2ex]{0pt}{5.5ex} Ogden 2nd order & $\sum\limits_{i=1}^{2} \frac{2\,\mu_i}{\alpha_i^2} \left(\bar{\lambda}_1^{\alpha_i} + \bar{\lambda}_2^{\alpha_i} + \bar{\lambda}_3^{\alpha_i} - 3\right)$ \\ 
\hline \rule[-2ex]{0pt}{5.5ex} Yeoh & $\sum\limits_{i=1}^{3} C_{i0}\,(\bar{I}_1-3)^i$ \\ 
\hline 
\end{tabular} 
\end{table}

A major drawback of the higher order models is their limited stability. Under uniaxial compression the polynomial model with the specific parameters obtained from the fit becomes unstable at a strain of only -17~\%. The Ogden model of second order has a large range of stability in compression (-34~\%), but reaches the stability limit in tension already at 184~\%.

\begin{figure}[hbtp]
\centering
\includegraphics[height=\picHeight]{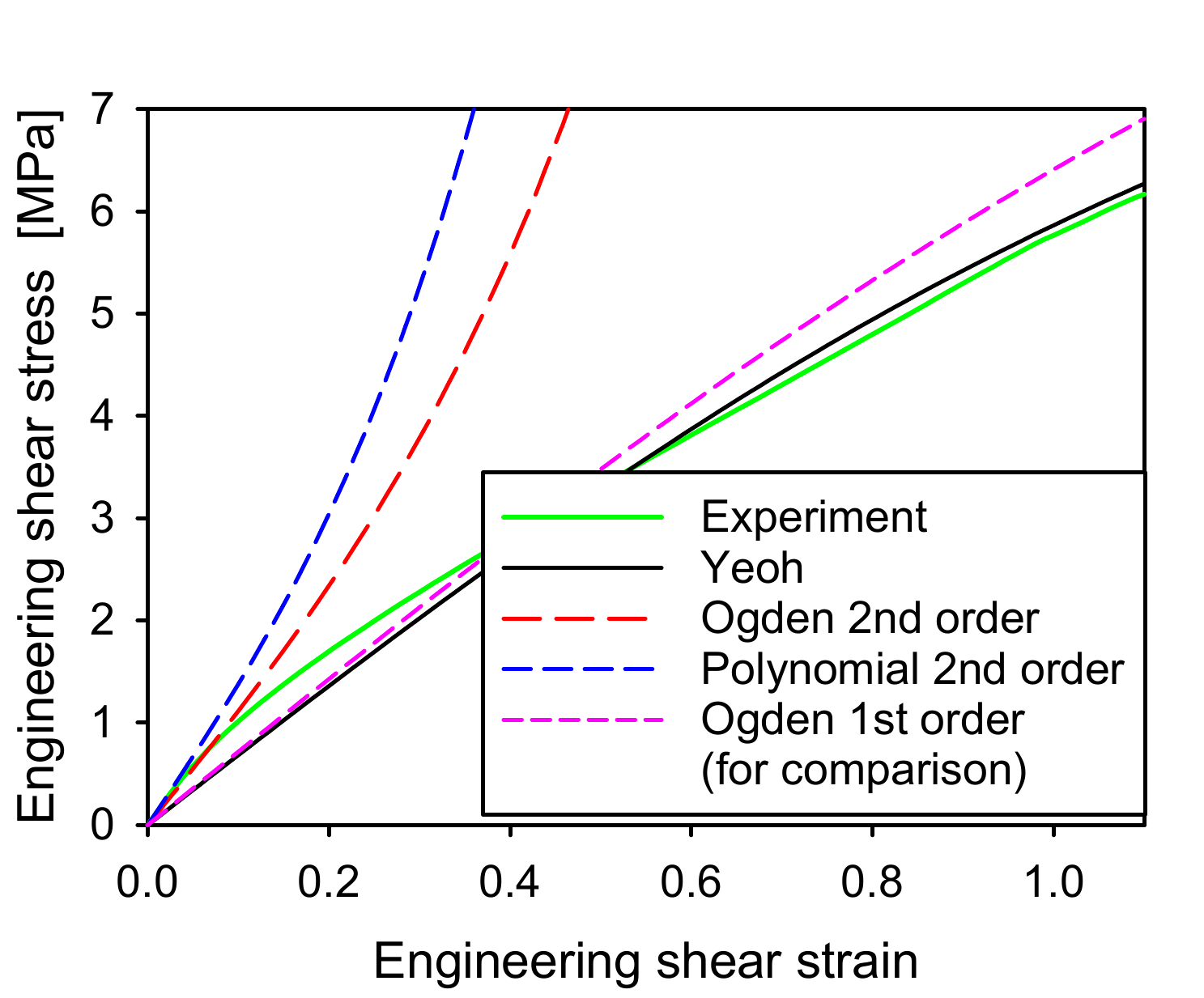}
\caption{Fit of lap shear test data by three conditionally stable hyperelastic models}
\label{fig:unstable-shear}
\end{figure}

A second disadvantage of the models of higher order which provide a good fit to the tensile test data is their extrapolation to other loadings. Figure \ref{fig:unstable-shear} shows the predictions of the same models for the lap shear tests. The result of the Ogden model of first order has been added to the figure for comparison, it gives a much better prediction of the shear behavior than the second order Ogden model.

The measurement curve in figure \ref{fig:unstable-shear} is approximately linear for strains larger than 0.2. The initial modulus is considerably higher than this slope, though, and decreases rapidly until a strain of about 0.1. This kind of behavior cannot easily be described by a polynomial model, but may be caught in a model which assumes a decay of a part of the modulus during the deformation up to a strain of 0.2.

In summary, section \ref{sec:established} has shown that all established hyperelastic models tested here are unable to provide an accurate fit of the experimental data of the considered adhesive. We made similar observations for other adhesive materials which are not shown in this paper.

\section{New viscoelastic-hyperelastic model}

\subsection{New hyperelastic potential}\label{sec:potential}

The aim of the new model is to close the gap between experimental observations and hyperelastic models at low strains without accepting the disadvantages of limited stability or poor extrapolation like in the Ogden model of second order. The basic idea is that a part of the stress decays exponentially with increasing deformation. We noticed later that this idea was already proposed by \citet{yeoh1993} who suggested to extend the well known Yeoh model, a reduced polynomial model of third order, by an additional exponential term. The strain energy density in this extended model of Yeoh is
\begin{align}
U_\mathrm{Yeoh,exp} &= \frac{A}{B} \,\left( 1 - e^{-B\,(\bar{I}_1-3)} \right) + U_\mathrm{Yeoh} \label{eq:yeoh-exp}\\
U_\mathrm{Yeoh} &= \sum\limits_{i=1}^{3} C_{i0}\,(\bar{I}_1-3)^i \label{eq:yeoh}
\end{align}
Here, the constants $C_{i0}$ denote the material parameters of the polynomial model, the parameter $A$ defines the magnitude of the contribution of the exponential term to the stresses, and the constant $B$ determines how fast the exponential term decays. 

We suggest a different approach to introduce the exponential decay into the hyperelastic potential. Instead of the additive contribution to the potential which results in an additive contribution to the stresses, the stresses are multiplied by a factor expressing the exponential decay. This approach can conveniently be applied to any existing hyperelastic model to introduce the exponential decay of a part of the initial stiffness. Let $U_\textrm{old}$ be a an arbitrary strain energy density function of a hyperelastic model not yet containing the exponential decay. Then, a new potential is defined by
\begin{equation}
U_\mathrm{new} = f(U_\mathrm{old})
\end{equation}
using a scalar function $f$ which needs yet to be determined. The stresses in hyperelastic models are determined by derivatives of the potential. According to the chain rule, any first derivative of the new potentials differs by a factor of the derivative of the old potential:
\begin{equation}
dU_\mathrm{new} = f'(U_\mathrm{old}) \,dU_\mathrm{old}
\end{equation}
The factor shall provide the exponential decay of a part of the stress at increasing deformation. This is achieved by choosing
\begin{equation}\label{eq:fstrich}
f'(U_\mathrm{old}) = 1 - c \left(1 - e^{-\frac{U_\mathrm{old}}{U_0}} \right)
\end{equation}
where $c$ is the decaying fraction of the initial stiffness and $U_0$ is a reference value of the old potential determining how fast the stiffness decays.

By integrating \eqref{eq:fstrich} the new potential with the desired property is obtained:
\begin{equation}\label{eq:f}
U_\mathrm{new} = f(U_\mathrm{old}) = (1 - c) \,U_\mathrm{old} + c \,U_0 \,\left(1 - e^{-\frac{U_\mathrm{old}}{U_0}} \right)
\end{equation}

This new approach can be applied to any existing hyperelastic model. In the following sections its application to an Ogden model of first order shall be considered in particular. The strain energy density of Ogden model consists of a deviatoric and a volumetric part:
\begin{align}
U_\mathrm{Ogden} &= U_\mathrm{dev} + U_\mathrm{vol} \label{eq:ogden}\\
U_\mathrm{dev} &= \frac{2\,\mu}{\alpha^2} \left(\bar{\lambda}_1^\alpha + \bar{\lambda}_2^\alpha + \bar{\lambda}_3^\alpha - 3\right) \label{eq:ogden-dev}\\
U_\mathrm{vol} &= \frac{K}{2} \,(J-1)^2 \label{eq:ogden-vol}
\end{align}
The constants $\mu,\alpha$ are defined in the notation employed in Abaqus so that $\mu$ has the meaning of the initial shear modulus. The original formulation of \citet{Ogden565} uses a factor $\mu/\alpha$ instead of $2\mu/\alpha^2$.
$K$ denotes the compression modulus. In a first step we formulate an incompressible model, thus we consider only the deviatoric part $U_\mathrm{dev}$ of the potential.

The new hyperelastic potential defined by equations \eqref{eq:f} to \eqref{eq:ogden-dev} was implemented in the commercial finite element code Abaqus using the user-subroutine UHYPER. The main difficulty of the implementation was due to different formulations used by the interface of Abaqus and the model: Abaqus requires the potential and its derivatives as a function of the strain invariants, while the Ogden model and consequently the new model are considering the potential as a function of the principal stretches. Nevertheless, an efficient implementation was created. In a test simulation of an adhesive joint the implementation of the new model demanded no significantly higher computational cost than the built-in implementation of the original Ogden model.

The parameter identification for the new hyperelastic model can be performed by fitting the stress-strain curve of the tensile test just like for other hyperelastic models. Experiment and model were evaluated at 100 time points, and the sum of the squared differences between measured and simulated stress was calculated. The minimization of this error value by sequential least squares programming lead to the optimized set of the four parameters ($\mu,\alpha,c,U_0$). Only the parameter $U_0$ appearing in the exponent in the function $f$ requires some special attention: If its initial value is chosen much too large or small, then the derivative of the error with respect to $U_0$ is very small and gradient based optimization algorithms will not find the optimal parameters. This risk can be avoided by optimizing the parameters for the original Ogden model ($c=0$) first. With these initial values for $\mu$ and $\alpha$ the value of the Ogden potential $U_\mathrm{Ogden}$ is calculated for some low value of strain in the uniaxial test, e.g.~10~\% strain. This gives a suitable initial value of $U_0$ for the parameter optimization.

\subsection{Compressibility}\label{compressible}

To extend the model to compressible material behavior, the volumetric part $U_\mathrm{vol}$ of the Ogden potential \eqref{eq:ogden} needs to be considered as well.
We apply the function $f$ modifying the potential only to the deviatoric part:
\begin{equation}\label{eq:uneu}
U = f(U_\mathrm{dev}) + U_\mathrm{vol}
\end{equation}
This leads to much less complicated derivatives of the potential than applying the function on the entire $U_\mathrm{Ogden}$, so that the implementation of the model is easier and the computational effort in finite elements is smaller. The choice is made for solely numerical reasons, because the available experimental data is not sufficient to decide whether a decay of stiffness under hydrostatic loads would be a more appropriate description of the adhesive material.

The compression modulus $K$ cannot be identified from the same tensile stress-strain curve as the other parameters of the hyperelastic model. In principal it should be possible to calculate the compression modulus from the Poisson's ratio $\nu$ obtained from a measurement of lateral contraction in the same tensile test in the limit of small strains:
\begin{equation}
K = \frac{2}{3} \frac{1+\nu}{1-2\,\nu} \,\mu
\end{equation}
However, the compression modulus is very sensitive to small changes of Poisson's ratio if the material is nearly incompressible ($\nu\approx 1/2$). Therefore, inevitable uncertainties and scatter of the measurement prohibit an accurate determination of the compression modulus from the tensile test.

Better suited to determine the compression modulus for nearly incompressible materials are tests with a higher trixiality of the stress state. Therefore, we employed two butt joint tensile tests to identify the compression modulus. Since the steel adherends have a much higher elastic modulus than the adhesive, the strains in the adherends are small and their lateral contraction may be neglected. For the linear elastic case, \citet{lindley1979} derived a closed form solution for the effective elastic modulus of a cylindrical, elastic layer of thickness $h$ and diameter $w$ between two rigid circular discs:
\begin{align}\label{eq:lindley}
\tilde{E} &= \begin{cases}
	2\,\mu + \frac{\lambda\mu}{\lambda+\mu} \,\left[ 1 + \frac{3}{8} \frac{\lambda\,(w/h)^2}{\lambda+\mu} \left( 1 - \frac{\mu\,(w/h)^2}{2(\lambda+\mu)+33\mu(w/h)^2/32}
	\right) \right]
	& \mathrm{if}\; w<w_2 \\
	\lambda + 2\,\mu - \frac{1}{15} \frac{\lambda^2}{\lambda+\mu}\frac{w_2}{w} \left( 8 - \frac{w_2}{w} \right)
	& \mathrm{if}\; w\ge w_2
\end{cases} \\
w_2 &= h \,\sqrt{\frac{64}{15}\frac{\lambda+\mu}{\mu}}
\end{align}

In this equation \eqref{eq:lindley} $\lambda$ does not denote a stretch but the Lam\'{e}'s first parameter. The ratio of the effective modulus $\tilde{E}$ considering the restriction of lateral contraction to the unconstrained elastic modulus $E$ depends on the adhesive layer aspect ratio $h/w$ and the ratio of compression to shear modulus $K/\mu$. Therefore, the ratio $K/\mu$ can be calculated if the stiffness of the butt joint has been measured for two different aspect ratios $h/w$.

\begin{figure}[hbtp]
\centering
\includegraphics[height=\picHeight]{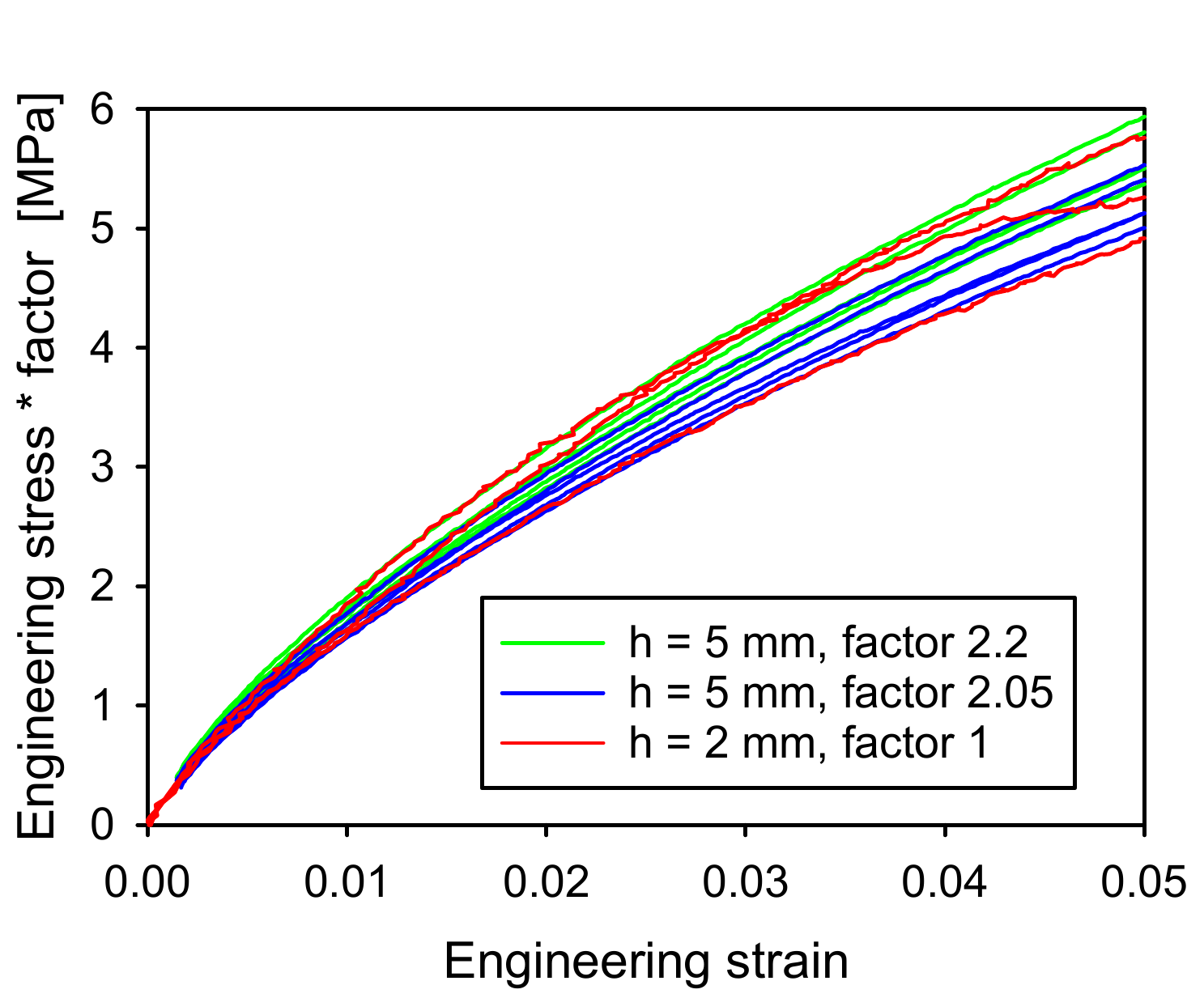}
\caption{Scaled stress-strain curves of butt joint tests}
\label{fig:butt-factor}
\end{figure}

The butt joint tests described in section \ref{sec:butt} were performed for one diameter $w$ and two different thicknesses $h$. The response of the specimen was non-linear, so that the solution of Lindley is not valid for large strains. However, the ratio of the stiffness of the specimens with thin adhesive layer to the specimens with thick adhesive layer was approximately constant for strains below 5~\%. This is shown in figure \ref{fig:butt-factor} where the red lines display the stress-strain curves of the specimens with thin adhessive layer. The specimens with thicker adhesive layer are less stiff, but if the stress is multiplied by a constant factor, these tests yield very similar curves. The blue curves origin from the tests with thick adhesive layer scaled by a factor 2.05, the blue curves are gained using a factor 2.2. It seems that the stiffness ratio between thin and thick adhesive layer is between 2.05 and 2.2. Using \eqref{eq:lindley} we calculate a compression modulus range $20.15\le K/\mu\le 24.9$ and select $K=22.4\mu$ which corresponds to a Poisson's ratio at small strains of 0.478.

\begin{figure}[hbtp]
\centering
\includegraphics[height=\picHeight]{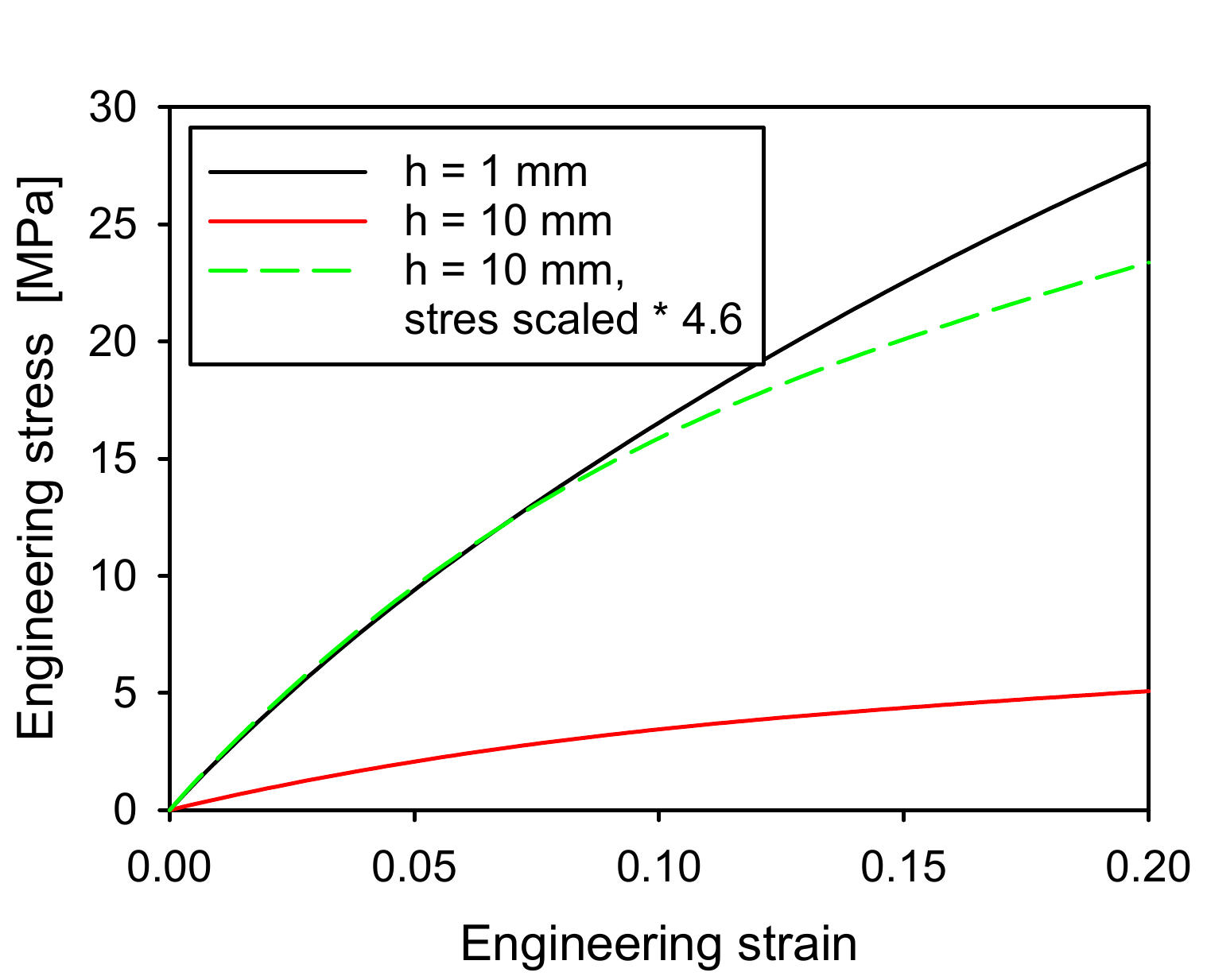}
\caption{Simulated stress-strain curves of butt joints with different adhesive thickness $h$}
\label{fig:butt-validity}
\end{figure}

Since equation \eqref{eq:lindley} strictly holds only for infinitely small strains and linear elastic behaviour, the validity of the evaluation approach for the non-linear material and small but finite strains must be checked. Therefore, finite element simulations of butt joint tests with two different values of adhesive layer thickness were performed using the viscoelastic-hyperelastic model and finite strains. The axial symmetric model of the adhesive layer used 8-node quadrilateral hybrid elements with linear pressure and reduced integration (CAX8RH). The element edge length was 0.25~mm. The displacement in load direction was prescribed for the nodes on the top and bottom of the adhesive layer. Figure \ref{fig:butt-validity} shows the engineering stress-strain curves obtained from the simulated butt joint tests. If the stresses evaluated for the specimen with the thicker adhesive layer are multiplied by a factor 4.6, the resulting curve fits well to the stress-strain curve of the thinner adhesive layer at strains below approximately 7~\% strain. The evaluation of this stiffness factor using \eqref{eq:lindley} yields a ratio of compression and shear modulus $K/\mu=22.3$ while the value used in the simulation was $K/\mu=22.4$.

\subsection{Viscoelastic-hyperelastic model}\label{visco}

The investigated adhesive exhibits a strain rate dependent behaviour at room temperature. This property was considered by combining the new hyperelastic model with linear viscoelasticity. Fortunately, it was not necessary to implement this viscoelastic-hyperelastic model as a user subroutine, since the employed finite element software (Abaqus) offers the capability to couple any user-defined hyperelastic model with linear viscoelasticity. The viscoelasticity model used is a generalized Maxwell model expressed as a Prony series with relaxation times $\tau_i$ and normalized coefficients $\bar{g}_i$:
\begin{equation}\label{eq:visco}
U_\mathrm{visco} = \left( 1 - \sum\limits_{i=1}^{N} \bar{g}_i \,\left(1 - e^{t/\tau_i} \right) \right) U
\end{equation}

\begin{figure}[hbtp]
\centering
\includegraphics[height=\picHeight]{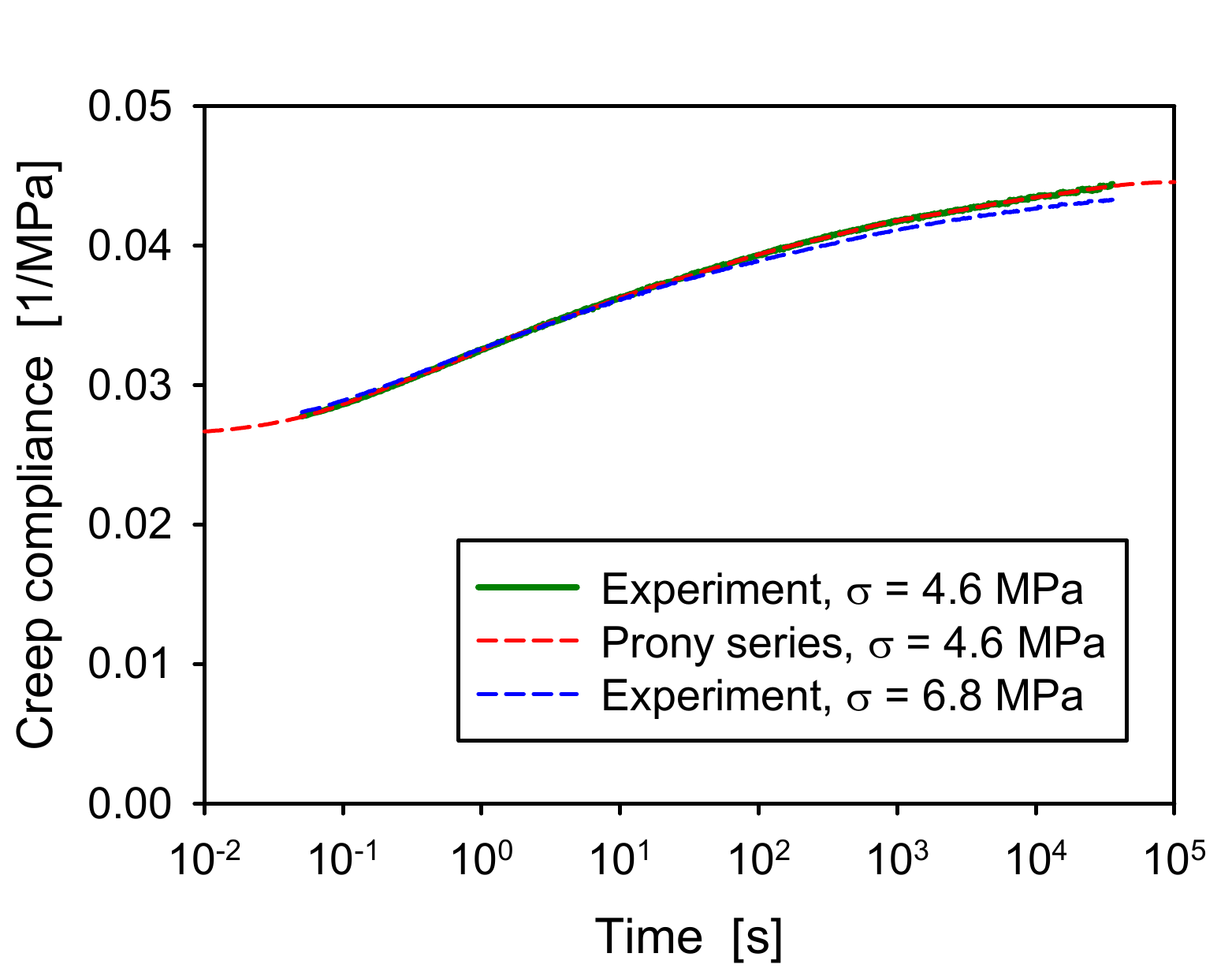}
\caption{Creep compliance at two load levels and fit by series}
\label{fig:creep}
\end{figure}

The parameters of the viscoelastic part of the model ($\tau_i,\bar{g}_i$) were identified using the creep compression tests. The tests results at the two different load levels showed only a small difference, so that their evaluation assuming linear viscoelasticity was justified. The creep compliance was evaluated as a function of time and fitted by a kind of Prony series with 7 terms (figure \ref{fig:creep}). A Laplace transformation was applied to obtain the corresponding relaxation function and the parameters of the Prony series \eqref{eq:visco}.

It should be noted that the creep test was not chosen because of preference but because of the availability of data in the project. In most cases a tensile relaxation test will be better suited for the identification of the viscoelastic parameters than the compressive creep test. Alternatively, the parameters can be identified using the results of tensile tests at several different strain rates. For the experimental data considered in this paper, the parameter identifications from the creep tests and from the tensile tests at different strain rates yielded very similar results.

Once the viscoelastic parameters have been identified from the creep test and the ratio between compression and shear modulus from the two butt joint tests, these parameters are kept constant. The remaining parameters of the hyperelastic model are now identified using the tensile test as described in section \ref{sec:potential}. The only difference from the determination of parameters of the model without viscoelasticity is that now the simulations of the tensile test consider the viscoelasticity.

\section{Results and discussion}

After the model has been formulated and the parameters identified, now the model predictions for those tests which have not been used for parameter identification shall be presented. The validation is split into two parts. First, we compare the results of the new hyperelastic model at a single strain rate to existing models. Next, we show the predictions of the new visoelastic-hyperelastic model at different strain rates.

\subsection{Hyperelastic models}

\begin{figure}[hbtp]
\centering
\includegraphics[height=\picHeight]{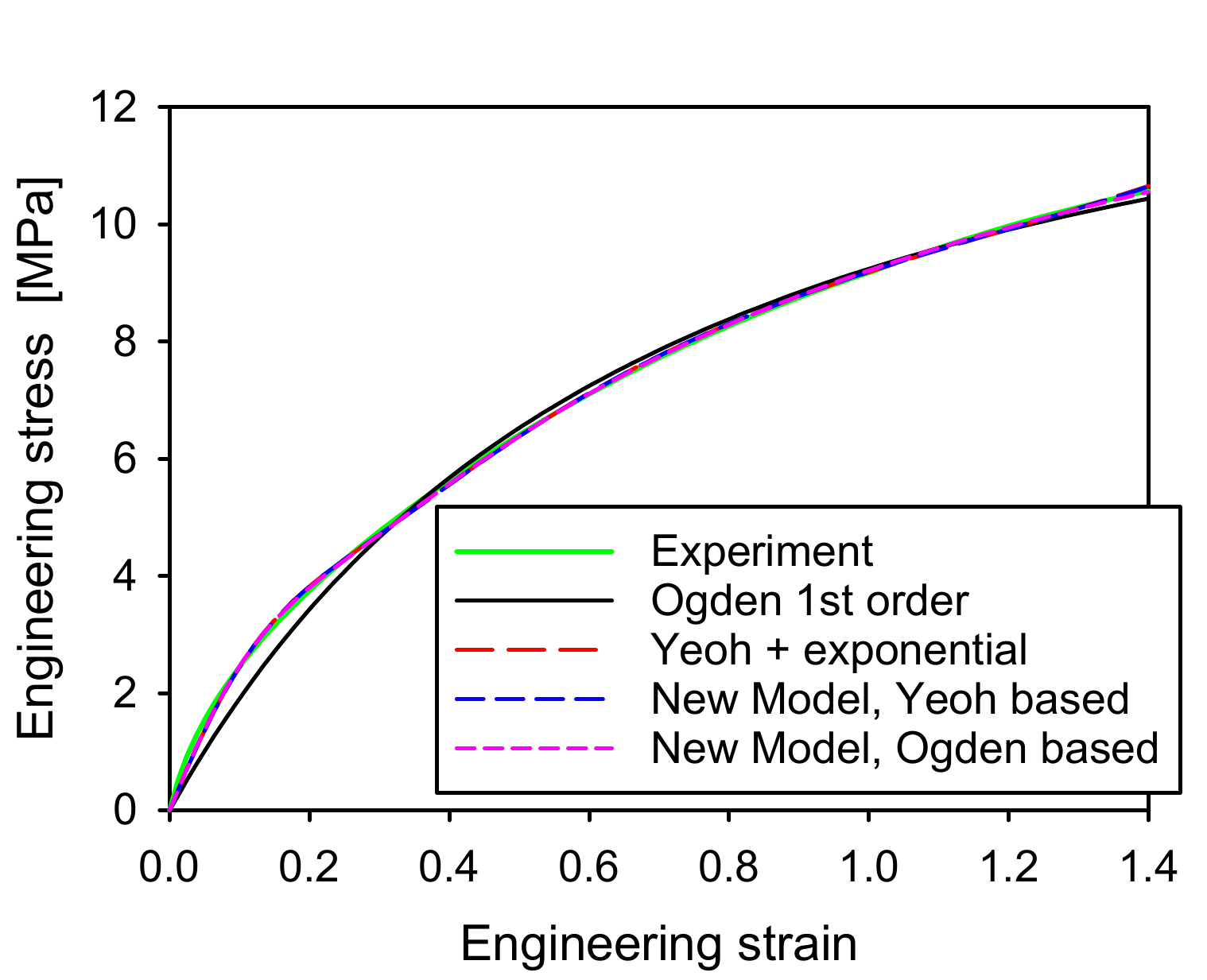}
\caption{Fit of models to uniaxial tensile test at $\dot{\varepsilon}=0.005 $s$^{-1}$}
\label{fig:extended-tension}
\end{figure}

Figure \ref{fig:extended-tension} displays a stress-strain curve of a tensile test and the results of four models fitted to this experimental data. The Ogden model of first order is shown as a reference for the other models. Those three models all include some exponential function to provide a decay of stiffness. This feature enables them to give a good description of the entire stress-strain curve, while the more simple Ogden model significantly deviates from the experiment concerning the initial stiffness.

The first of the three models using the exponential function is the model of \citet{yeoh1993}. The second model uses our new approach to introduce the exponential decay, but is using on the same basic Yeoh model \citep[eq.~\eqref{eq:yeoh},][]{yeoh1990} as its basic potential ($U_\mathrm{old}=U_\mathrm{Yeoh}$). The third model also uses the new approach and is based on the Ogden potential of fist order ($U_\mathrm{old}=U_\mathrm{Ogden}$). The values of the parameters identified for the three models are given in table \ref{tab:parameters}.

\begin{table}[htbp]
\centering
\caption{Model parameters \label{tab:parameters}}
\begin{tabular}{ccccc}
\hline \multicolumn{5}{l}{Ogden 1st order} \\ 
\hline $\mu$ & $\alpha$ &&& \\ 
\hline 7.17~MPa & 0.988 &&& \\
\hline \multicolumn{5}{l}{Yeoh with exponential term} \\ 
\hline $C_{10}$ & $C_{20}$ & $C_{30}$ & $A$ & $B$  \\ 
\hline 3.23~MPa & -0.196~MPa & -0.0147~MPa & 1.66~MPa & 9.67  \\ 
\hline \multicolumn{5}{l}{New model based on Yeoh} \\ 
\hline $C_{10}$ & $C_{20}$ & $C_{30}$ & $c$ & $U_0$  \\ 
\hline 4.89~MPa & -0.290~MPa & 0.0212~MPa & 0.340 & 0.511  \\ 
\hline \multicolumn{5}{l}{New model based on Ogden} \\ 
\hline $\mu$ & $\alpha$ & $c$ & $U_0$ & \\ 
\hline 10.1~MPa & 1.13 & 0.317 & 0.453 &  \\ 
\hline 
\end{tabular} 
\end{table}

\begin{figure}[hbtp]
\centering
\includegraphics[height=\picHeight]{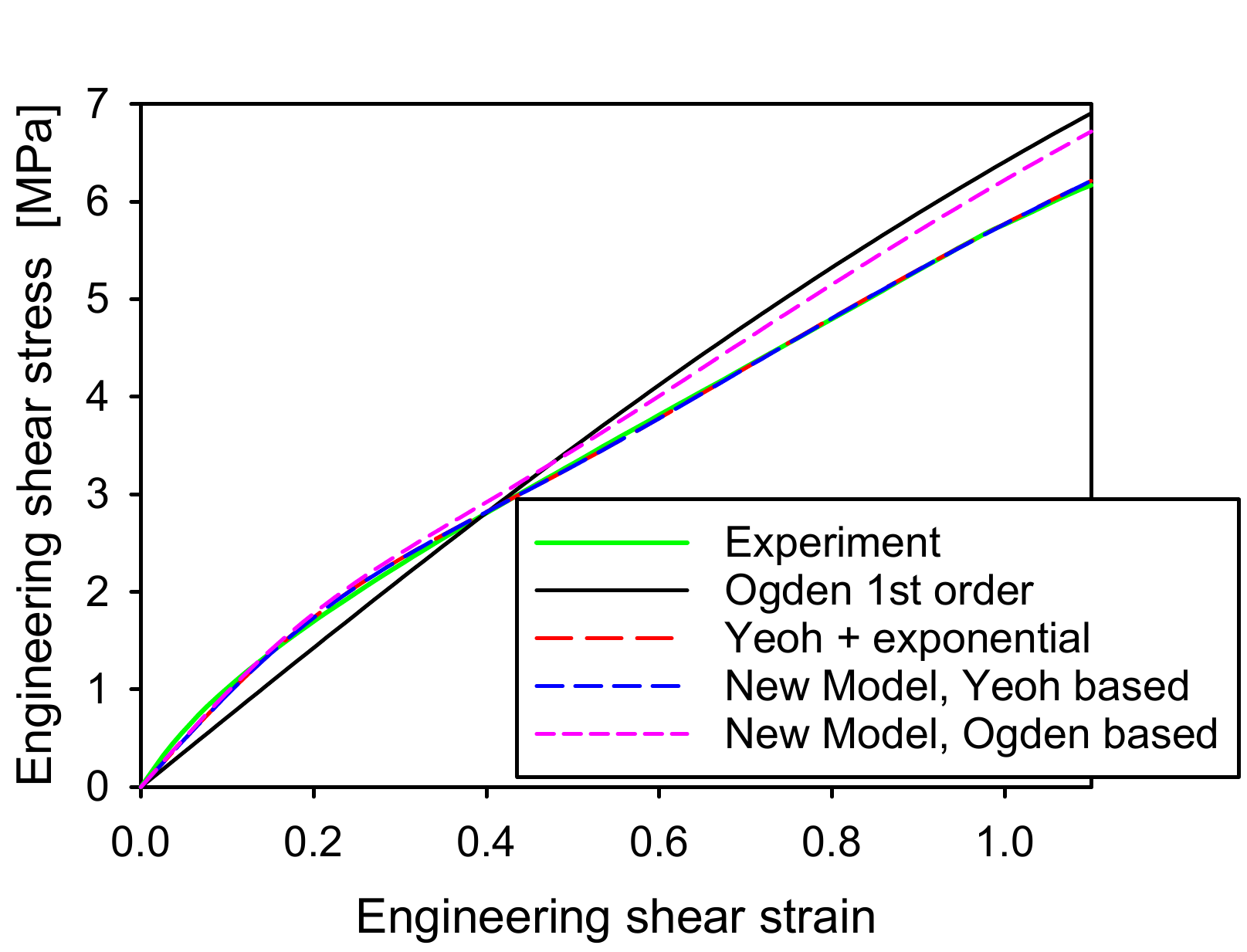}
\caption{Prediction of models for lap shear test at $\dot{\gamma}=0.01 $s$^{-1}$}
\label{fig:extended-shear}
\end{figure}

Lap shear tests and compressive tests have been chosen to test the prediction of the models under different loading conditions. Since the hyperelastic models without viscoelasticity cannot consider the strain rate dependence of the material behavior, care was taken to perform the tests at similar strain rates. Figure \ref{fig:extended-shear} displays the results for the shear loading. The measurement curve shows a bend at the start of the test and an approximately linear behavior afterwards. The Ogden model fails to capture this feature of the curve even qualitatively. The models containing the exponential functions, however, predict the bend in the curve very well. The Yeoh model with exponential term makes an excellent prediction up to 110~\% shear strain which was close before crack initiation in the lap shear specimen. The model created by the new approach starting from the Yeoh model yields the same result. The Ogden model extended by the new approach overestimates the stiffness at large strains. Still, the relative error of the stress prediction of this model at 110~\% shear strain is only 9~\%.

\begin{figure}[hbtp]
\centering
\includegraphics[height=\picHeight]{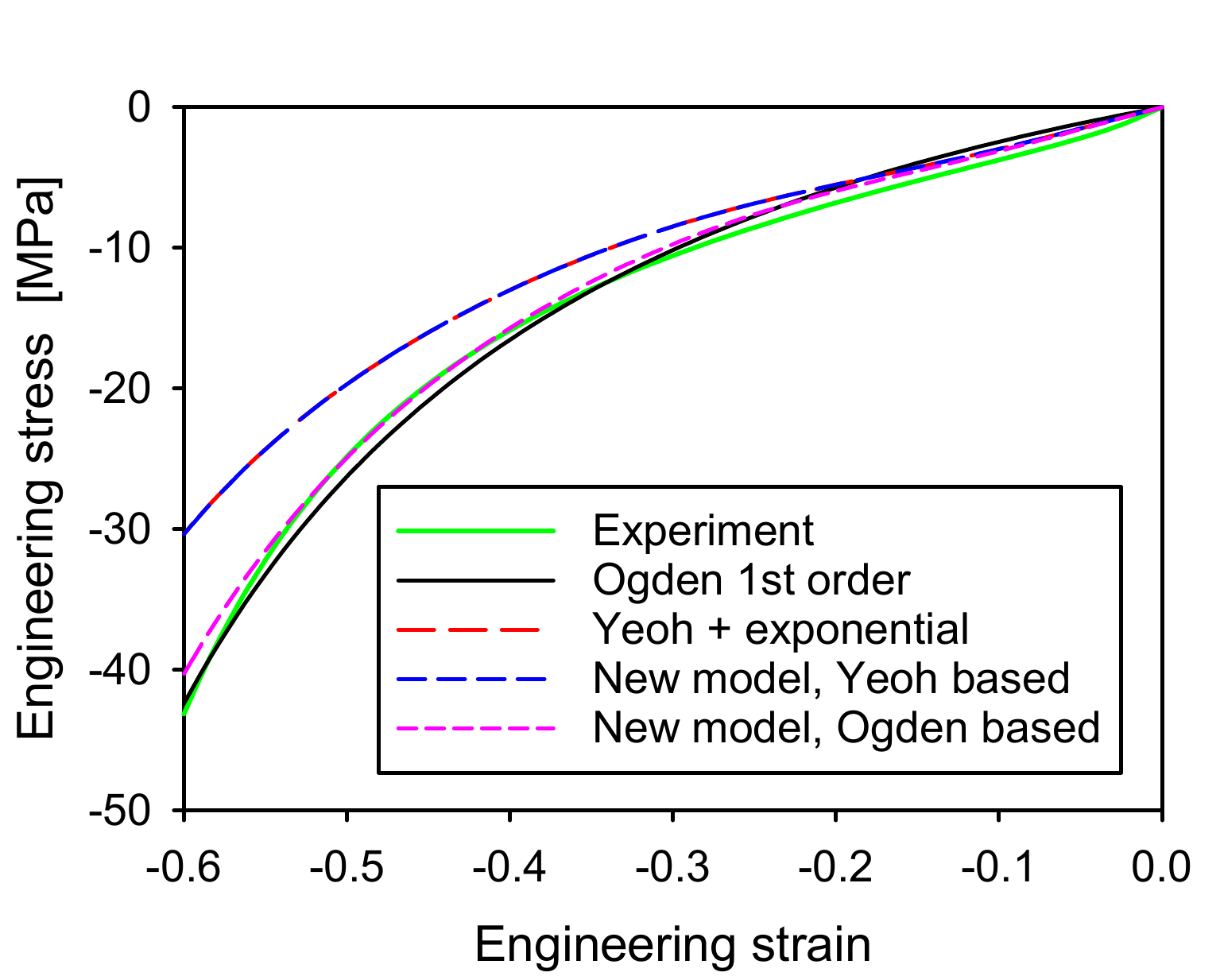}
\caption{Prediction of models for uniaxial compression test at $\dot{\varepsilon}=0.005 $s$^{-1}$}
\label{fig:extended-compression}
\end{figure}

In the simulation of the compressive test in figure \ref{fig:extended-compression} the new model based on the Ogden potential predicts the experiment significantly better than the model of \citet{yeoh1993} or the new model based on the Yeoh model.

In the simulations of all three experiments the results of the new model based on the Yeoh potential are nearly indistinguishable from the results of the Yeoh model with the additional exponential term, although the model equations are different. The reason is that for the parameters fitting to the tensile test (table \ref{tab:parameters}) the exponential term in the potential decays almost completely before the higher order terms in the Yeoh potential become significant, i.e.
\begin{equation}
e^{-\frac{1}{U_0}\sum\limits_{i=1}^{3} C_{i0}\,(\bar{I}_1-3)^i} \approx e^{-\frac{1}{U_0}\,C_{10}\,(\bar{I}_1-3)}
\end{equation}
With this approximation the new model based on the Yeoh potential simplifies to
\begin{equation}
U_\mathrm{new} \approx \sum\limits_{i=1}^{3} (1 - c) \,C_{i0}\,(\bar{I}_1-3)^i + c \,U_0 \,\left( 1 - e^{-\frac{C_{10}}{U_0}\,(\bar{I}_1-3)} \right)
\end{equation}
The comparison to the Yeoh model with the additional exponential term
\begin{equation}
U_\mathrm{Yeoh,exp} = \sum\limits_{i=1}^{3} (1 - c) \,C_{i0}\,(\bar{I}_1-3)^i + \frac{A}{B} \,\left( 1 - e^{-B\,(\bar{I}_1-3)} \right)
\end{equation}
shows that the two potentials are approximately equal for a certain choice of parameters, e.g.~$c\,U_0=A/B$. A check with the parameters of table \ref{tab:parameters} confirms that the parameter optimizations found two equivalent sets of parameters.

Considering the quality of the model fit to the tensile tests and the results of the model validation in lap shear and compressive tests, it is evident that the idea of an exponential decay of a part of the stiffness leads to a significant improvement of the description of materials like the investigated adhesive. The validation on the experimental data presented in this paper does not allow to show unambiguously if a model based on the Yeoh potential or the Ogden potential simulates the real material behavior better than the other. Anyway, it seems likely that such a result would depend on the specific investigated material.

Keeping in mind that there is no universally favourable hyperelastic model, it is desirable to be able to apply the modification with a decaying stiffness to any hyperelastic potential to improve the model for low strains. The approach suggested in this paper \eqref{eq:f} is constructed to be applicable to any hyperelastic model. The additional exponential term \eqref{eq:yeoh-exp} proposed by \citet{yeoh1993} can also be added to any hyperelastic model. According to our best knowledge, so far it has been applied to the models of \citet{yeoh1990} and \citet{gent1996} \citep[see][]{yeoh1993,yeoh1997}.

Let us now compare the approaches of Yeoh and of the current paper to extend arbitrary hyperelastic models:
\begin{align}
U_\mathrm{Yeoh,new} &= U_\mathrm{old} + \frac{A}{B} \,\left( 1 - e^{-B\,(\bar{I}_1-3)} \right) \label{eq:compyeoh}\\
U_\mathrm{new} &= (1-c) \,U_\mathrm{old} + c \,U_0 \,\left(1 - e^{-\frac{U\mathrm{old}}{U_0}} \right) \label{eq:compus}
\end{align}
The factor $1-c$ in front of $U_\mathrm{old}$ does not lead to a different model, because it is equivalent to a change of some model parameters (e.g.~$C_{ij}$ in polynomial models or $\mu_i$ in Ogden models) by this factor. If the coefficients in front of the large brackets are chosen so that $A/B=c\,U_0$, then the only remaining difference is the exponent. In our model \eqref{eq:compus} the exponent consists of the original hyperelastic potential scaled by some parameter. In the approach of Yeoh \eqref{eq:compyeoh} it is chosen to be $\bar{I}_1-3$ scaled by some parameter, which is the potential of the neo-Hooke model. This means that the approach of Yeoh combines to kinds of dependencies on the strain invariants or on the principal stretches: the basic model $U_\mathrm{old}$ depends on the invariants in a certain way, and the exponential term has the simple dependence on $\bar{I}_1$ of the neo-Hooke model. In this regard our approach is a more straightforward model extension, since it does not change the way the invariants $\bar{I}_1,\bar{I}_2$ or stretches enter the potential.

In many cases the difference between the two approaches will be negligibly small like it was shown in figures \ref{fig:extended-tension} to \ref{fig:extended-compression}. The exponential term is responsible for a decay of stiffness at low to moderate strains, it is irrelevant for larger strains. In the limit of small strains the hyperelastic models usually converge to a behavior of neo-Hookean type. This has been numerically tested using the test data of our investigated adhesive for the extensions of the Mooney-Rivlin model and of the first order Ogden model. In both cases, no significant differences between the two approaches of model extension were visible in the simulations of uniaxial and shear tests. For the extensions of the Ogden model we will now consider the differences analytically.

The two potentials $U_\mathrm{new}$ and $U_\mathrm{Yeoh,new}$ can describe the same material behavior, if the exponents can be made equal by an appropriate choice of the parameters $B$ and $U_0$. This means that the original potential $U_\mathrm{old}$ needs to be proportional to $\bar{I}_1-3$, and the deviation from this proportionality determines the amount of difference between the two approaches. If we consider the Ogden model of first order \eqref{eq:ogden-dev} as the basis of the model extension, consider uniaxial loading with $\lambda_1=\lambda_u,\lambda_2=\lambda_3=\lambda_u^{1/2}$ and make use of the definition of the invariants \eqref{eq:invariants}, we obtain a relation between $U_\mathrm{old}$ and $\bar{I}_1-3$. A Taylor series expansion around $\lambda_u=1$ gives
\begin{equation}
\frac{U_\mathrm{old}}{\bar{I}_1-3} = \frac{\mu}{2} \,\left[1 + \frac{\alpha-2}{6\sqrt{3}} (\bar{I}_1-3)^{1/2} + \frac{3\,\alpha^2-4\,\alpha-4}{144} (\bar{I}_1-3) + \mathcal{O}\left((\bar{I}_1-3)^{3/2}\right) \right]
\end{equation}
The terms in the rectangular brackets containing $\bar{I}_1-3$ determine the amount of deviation from the proportionality. If we set the model parameter $\alpha=1.13$ and assume that the decay of stiffness happens in the range $\lambda<1.2$, then the magnitude of the term of order $1/2$ is less than 3~\% and the term of order $1$ less than 0.4~\%. In the same way the case of shear loading can be considered and results in the relation
\begin{equation}
\frac{U_\mathrm{old}}{\bar{I}_1-3} = \frac{\mu}{2} \,\left[1 + \frac{\alpha^2-4}{48} (\bar{I}_1-3) + \mathcal{O}\left((\bar{I}_1-3)^{3/2}\right) \right]
\end{equation}
which deviates even less from proportionality than the results for uniaxial stresses if $\alpha<2$.

\subsection{Viscoelastic-hyperelastic model}

The combination of the hyperelastic models discussed in the previous subsection with viscoelasticity is supposed to be able to predict the material behaviour at different strain rates. Now the model consisting of linear viscoelasticity, the Ogden model of first order and its extension according to the new approach defined in section \ref{sec:potential} will be validated.

The model parameters were identified using the relaxation test, the stiffness ratio comparing butt joint tests at two values of adhesive thickness, and a tensile test at one strain rate (0.05~s$^{-1}$). From the tests listed in section \ref{sec:experiments} this leaves the tensile tests at the other strain rates, the lap shear tests, the compression tests, and not yet used data from the butt joint tests to validate the model.

The identification of the parameters of viscoelasticity was described in section \ref{visco}, the identification of the ratio between compression and shear modulus from the butt joint tests in section \ref{compressible}. To identify the missing parameters of the hyperelastic model, simulations of the tensile tests were performed using the complete material model containing both viscoelasticity and hyperelasticity. Due to the homogeneity of the tensile test this simulation could simply be performed using a single finite element. The parameters of hyperelasticity were optimized, until a good match between simulation and measurement of the tensile test was reached. This parameter identification does not assume that the material behaviour at the specific strain rate of the tensile test is purely elastic without visoelasticity.

In the following subsections the results of the validation will be discussed. In order keep the size of the paper within limits, the simulated stress-strain curves have been displayed together with the results of the experiments in section \ref{sec:experiments}.  The measurement curves of all individual specimens are shown in different shades of green to give an impression of the experimental scatter. The blue curve is the result of a simulation using the new model. As a reference the result of a usual Ogden model of first order combined with linear viscoelasticity is shown as red curve.

The selection of the types of experiments used in the validation was made for reasons of simplicity of tests (uniaxial tension and compression) and relevance for adhesive joints (lap shear and butt joint tests). It should be noted that this selection does not cover all possible stress states. For example, equi-biaxial tests are considered an important test case for rubber-like materials in general \citep[e.g.][]{marckmann2006}. In the special case of adhesive joints the shear dominated stresses in lap shear tests and the high restriction of lateral contraction in butt joint tests are representative for the most important load cases in applications.

\subsubsection{Tensile test}\label{sec:val-tensile}

The results of the tensile tests not used for parameter identification are displayed in figures \ref{fig:tensile-0p005}, \ref{fig:tensile-0p05}, and \ref{fig:tensile-5}. The strain rate dependence is predicted well by the model. The new model but not the basic Ogden model succeeds to predict the high stiffness at low strains.

The tensile test at a strain rate of 0.05~1/s was used for parameter identification, so that the agreement between experiment and simulation in figure \ref{fig:tensile-0p05} shows the success of the parameter fit but should not be considered as a validation of a model prediction.

\subsubsection{Lap shear test}\label{sec:val-shear}

The results of lap shear tests at three different strain rates are shown in figures \ref{fig:shear-0p01} to \ref{fig:shear-5}. Both strain rate dependence and the bend in the beginning of the curve are predicted well by the new model. At shear strains larger than 50~\% the models overestimate the stress slightly. This overestimation is not a special feature of the new model, but in that strain range the new model and the basic Ogden model predict the same stress-strain curve.

It should be noted that no test with a shear load was used for parameter identification, but only uniaxial test data.

\subsubsection{Compression test}

\begin{figure}[hbtp]
\centering
\includegraphics[height=\picHeight]{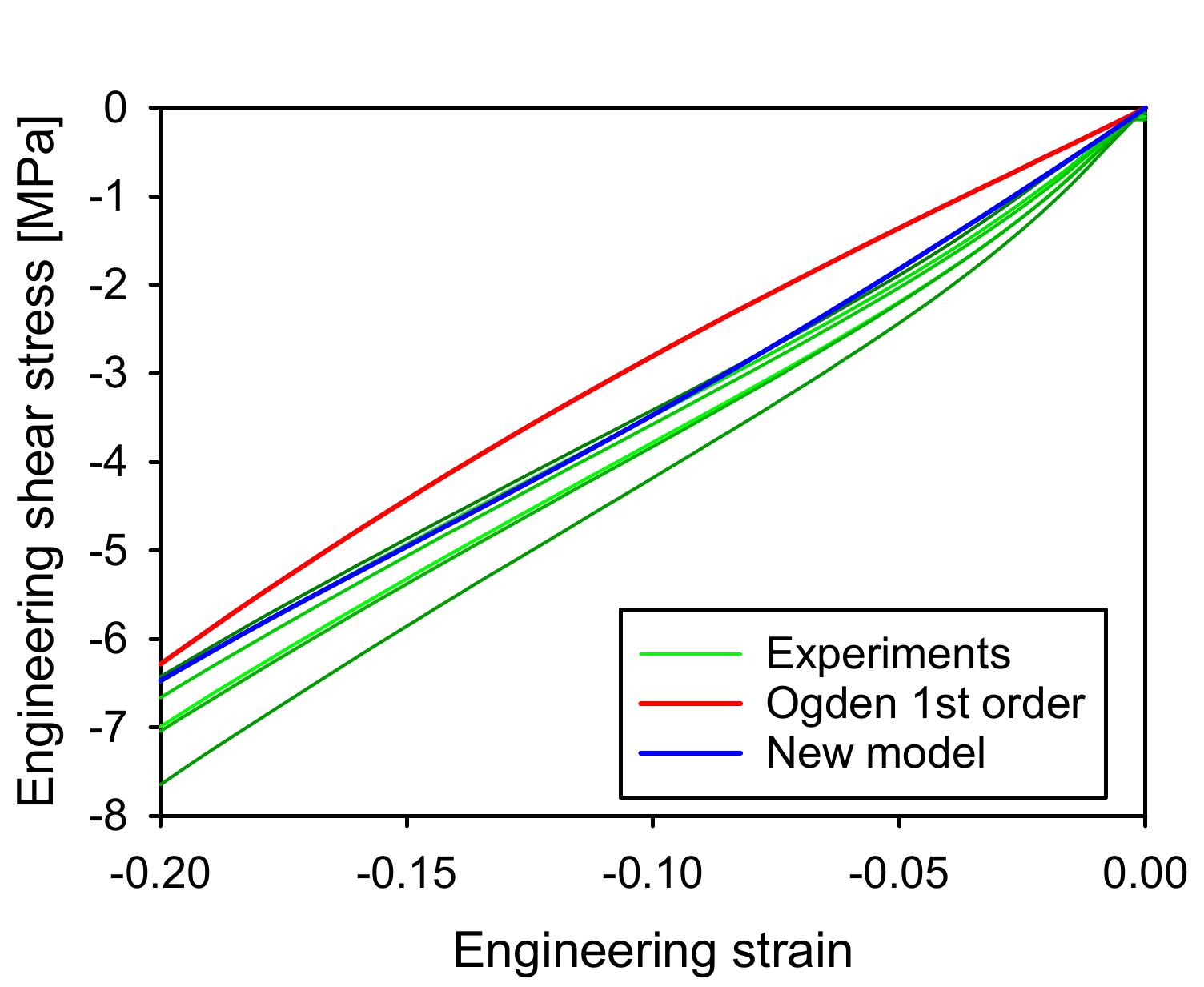}
\caption{Stress-strain curve of uniaxial compression test at $\dot{\varepsilon}=0.05~$s$^{-1}$ and simulation, detail at low strains}
\label{fig:compression-0p05-start}
\end{figure}

The results of uniaxial compressive tests at three different strain rates are shown in figures \ref{fig:compression-0p0005} to \ref{fig:compression-0p05}. The model predicts the strain rate dependence satisfyingly. Figure \ref{fig:compression-0p05-start} shows the beginning of the test at 0.05~s$^{-1}$ up to 20~\% shear strain so that the improvement by the new model at low strains is better visible.

\subsubsection{Butt joint test}

\begin{figure}[hbtp]
\centering
\includegraphics[height=\picHeight]{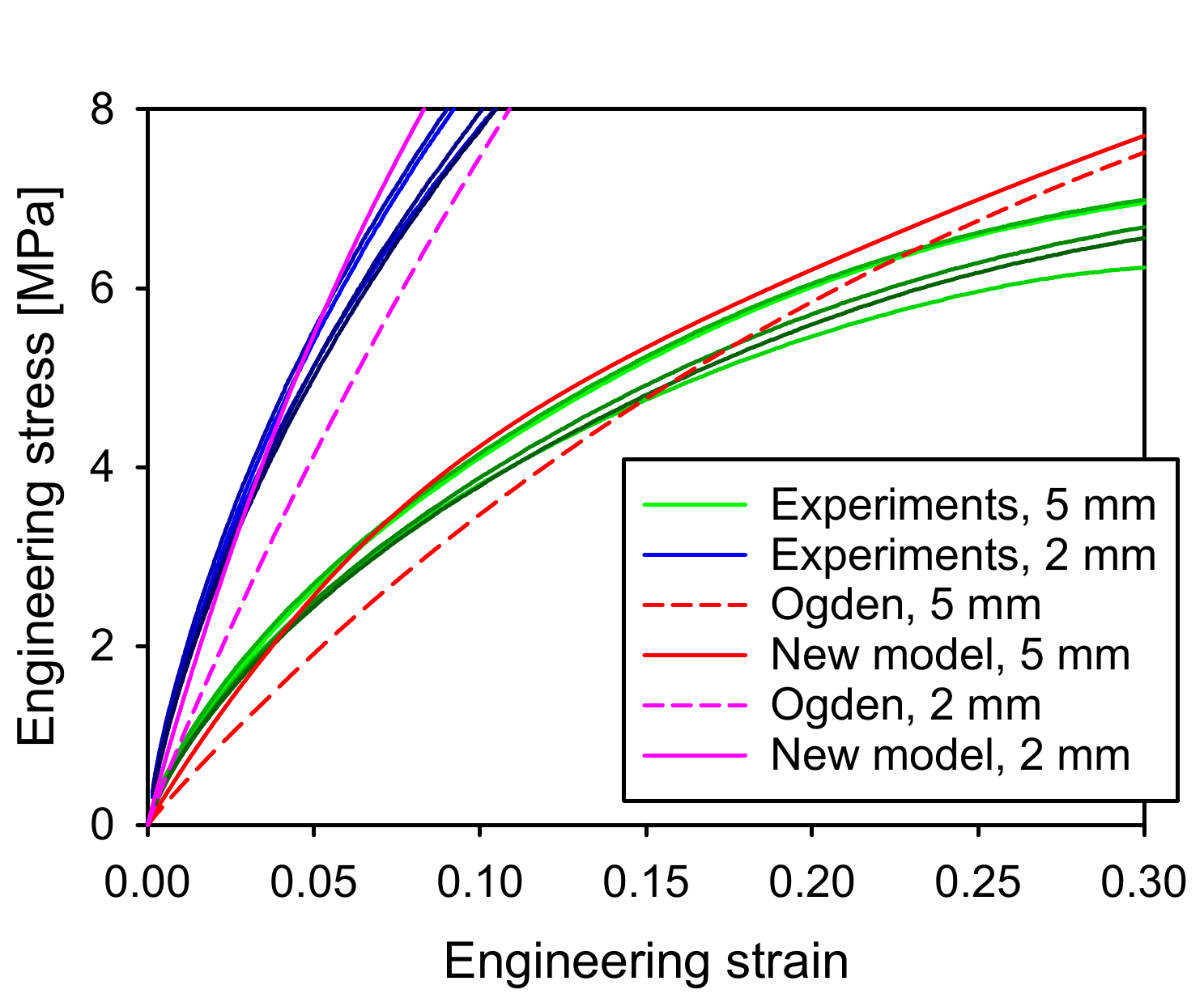}
\caption{Stress-strain curve of butt joint tensile test at $\dot{\varepsilon}=0.005~$s$^{-1}$ and simulation for adhesive layer thickness 2 and 5~mm}
\label{fig:butt}
\end{figure}

The butt joint tests at both 2~mm and 5~mm adhesive layer thickness have been employed in the parameter identification. However, only the stiffness ratio comparing both tests was used, so that the absolute value of the stiffness as well as the shape of the curve can still be used for validation. The results are displayed in figure \ref{fig:butt}.

The simulation of the butt joint tests used an axial symmetric model of the adhesive layer using 8-node quadrilateral hybrid elements with linear pressure and reduced integration (CAX8RH). The element edge length was 0.25~mm. The displacement in load direction was prescribed for the nodes on the top and bottom of the adhesive layer.

\section{Conclusions}

The material considered in the experimental part of this paper is a typical polyurethane adhesive which is used in the automotive industry. While it exhibits a hyperelastic mechanical behaviour, the well established hyperelastic models available in commercial finite element software fail to describe it accurately. The measured stiffness of the material at low strains is about 50~\% higher than the model prediction. This gap between experiment and simulation is closed by the proposed model as well as by the approach of \citet{yeoh1993}.

The new model was implemented via user subroutine in a commercial finite element code. No significant increase of the overall numerical cost of the simulation due to the more complex model was noticed. Two additional model parameters are required, but their identification requires no additional experiments and caused no difficulties. Therefore, the use of the extended model can be recommended for all those hyperelastic materials which show a similar deviation from the standard models.

The model was combined with linear viscoelasticity and succeeded to predict the rate dependence of several tests. It should be noted that this rate dependent data was not used in the parameter identification. A uniaxial tensile test at one constant strain rate, one creep test and two butt joint tests were sufficient to obtain the model parameters. The calibrated model predicted tensile, compressive and shear tests at several strain rates well. From an application point of view it is very appealing to get this quality of prediction based on such a low number of experiments for parameter identification. Future research will show whether this low experimental effort of the parameter identification is specific to the investigated adhesive or sufficient for a broad class of materials.

\section*{Acknowledgements}

The IGF project 17276~N "Numerische Modellierung und Kennwert\-er\-mitt\-lung f{\"u}r das Versagensverhalten von hyperelastischen Klebverbindungen" of the research association Forschungsvereinigung Stahlanwendung e.V. (FOSTA), Sohnstra\ss{}e 65, D-40237 D{\"u}sseldorf was funded by the AiF under the program for the promotion of joint industrial research and development (IGF) by the Federal Ministry of Economics and Energy based on a decision of the German Bundestag. The authors thank Georg Schwarzkopf of Paderborn University who performed the tests on butt joints.

\section*{References}

\bibliography{mybibfile}

\begin{thebibliography}{29}
\expandafter\ifx\csname natexlab\endcsname\relax\def\natexlab#1{#1}\fi
\providecommand{\url}[1]{\texttt{#1}}
\providecommand{\href}[2]{#2}
\providecommand{\path}[1]{#1}
\providecommand{\DOIprefix}{doi:}
\providecommand{\ArXivprefix}{arXiv:}
\providecommand{\URLprefix}{URL: }
\providecommand{\Pubmedprefix}{pmid:}
\providecommand{\doi}[1]{\href{http://dx.doi.org/#1}{\path{#1}}}
\providecommand{\Pubmed}[1]{\href{pmid:#1}{\path{#1}}}
\providecommand{\bibinfo}[2]{#2}
\ifx\xfnm\undefined \def\xfnm[#1]{\unskip,\space#1}\fi
\bibitem[{Arruda and Boyce(1993)}]{arruda1993three}
\bibinfo{author}{Arruda\xfnm[ E.M.]}, \bibinfo{author}{Boyce\xfnm[ M.C.]}.
\newblock \bibinfo{title}{A three-dimensional constitutive model for the large
  stretch behavior of rubber elastic materials}.
\newblock \bibinfo{journal}{Journal of the Mechanics and Physics of Solids}
  \bibinfo{year}{1993};\bibinfo{volume}{41}(\bibinfo{number}{2}):\bibinfo{pages}{389--412}.
\bibitem[{Balabaev and Khazanovich(2009)}]{balabaev2009extension}
\bibinfo{author}{Balabaev\xfnm[ N.]}, \bibinfo{author}{Khazanovich\xfnm[ T.]}.
\newblock \bibinfo{title}{Extension of chains composed of freely joined elastic
  segments}.
\newblock \bibinfo{journal}{Russian Journal of Physical Chemistry B}
  \bibinfo{year}{2009};\bibinfo{volume}{3}(\bibinfo{number}{2}):\bibinfo{pages}{242--246}.
\bibitem[{Boyce(1996)}]{boyce1996}
\bibinfo{author}{Boyce\xfnm[ M.C.]}.
\newblock \bibinfo{title}{Direct comparison of the {G}ent and the
  {A}rruda-{B}oyce constitutive models of rubber elasticity}.
\newblock \bibinfo{journal}{Rubber Chemistry and Technology}
  \bibinfo{year}{1996};\bibinfo{volume}{69}(\bibinfo{number}{5}):\bibinfo{pages}{781--785}.
\newblock \DOIprefix\doi{10.5254/1.3538401}.
\bibitem[{Chagnon et~al.(2015)Chagnon, Rebouah and Favier}]{Chagnon2015}
\bibinfo{author}{Chagnon\xfnm[ G.]}, \bibinfo{author}{Rebouah\xfnm[ M.]},
  \bibinfo{author}{Favier\xfnm[ D.]}.
\newblock \bibinfo{title}{Hyperelastic energy densities for soft biological
  tissues: A review}.
\newblock \bibinfo{journal}{Journal of Elasticity}
  \bibinfo{year}{2015};\bibinfo{volume}{120}(\bibinfo{number}{2}):\bibinfo{pages}{129--160}.
\newblock \DOIprefix\doi{10.1007/s10659-014-9508-z}.
\bibitem[{{Dassault Systemes}(2014)}]{abaqus}
\bibinfo{author}{{Dassault Systemes}\xfnm[]}.
\newblock \bibinfo{title}{Abaqus 6.14}; \bibinfo{year}{2014}. .
\bibitem[{Gent(1996)}]{gent1996}
\bibinfo{author}{Gent\xfnm[ A.N.]}.
\newblock \bibinfo{title}{A new constitutive relation for rubber}.
\newblock \bibinfo{journal}{Rubber Chemistry and Technology}
  \bibinfo{year}{1996};\bibinfo{volume}{69}(\bibinfo{number}{1}):\bibinfo{pages}{59--61}.
\newblock \DOIprefix\doi{10.5254/1.3538357}.
\bibitem[{G{\"o}ktepe and Miehe(2005)}]{goktepe2005micro}
\bibinfo{author}{G{\"o}ktepe\xfnm[ S.]}, \bibinfo{author}{Miehe\xfnm[ C.]}.
\newblock \bibinfo{title}{A micro--macro approach to rubber-like materials.
  {Part III}: The micro-sphere model of anisotropic mullins-type damage}.
\newblock \bibinfo{journal}{Journal of the Mechanics and Physics of Solids}
  \bibinfo{year}{2005};\bibinfo{volume}{53}(\bibinfo{number}{10}):\bibinfo{pages}{2259--2283}.
\bibitem[{Itskov and Knyazeva(2016)}]{itskov2016rubber}
\bibinfo{author}{Itskov\xfnm[ M.]}, \bibinfo{author}{Knyazeva\xfnm[ A.]}.
\newblock \bibinfo{title}{A rubber elasticity and softening model based on
  chain length statistics}.
\newblock \bibinfo{journal}{International Journal of Solids and Structures}
  \bibinfo{year}{2016};\bibinfo{volume}{80}:\bibinfo{pages}{512--519}.
\bibitem[{Kuhn(1936)}]{kuhn1936beziehungen}
\bibinfo{author}{Kuhn\xfnm[ W.]}.
\newblock \bibinfo{title}{Beziehungen zwischen {M}olek{\"u}lgr{\"o}{\ss}e,
  statistischer {M}olek{\"u}lgestalt und elastischen {E}igenschaften
  hochpolymerer {S}toffe}.
\newblock \bibinfo{journal}{Colloid \& Polymer Science}
  \bibinfo{year}{1936};\bibinfo{volume}{76}(\bibinfo{number}{3}):\bibinfo{pages}{258--271}.
\bibitem[{Lindley(1979)}]{lindley1979}
\bibinfo{author}{Lindley\xfnm[ P.B.]}.
\newblock \bibinfo{title}{Compression moduli for blocks of soft elastic
  material bonded to rigid end plates}.
\newblock \bibinfo{journal}{The Journal of Strain Analysis for Engineering
  Design}
  \bibinfo{year}{1979};\bibinfo{volume}{14}(\bibinfo{number}{1}):\bibinfo{pages}{11--16}.
\newblock \DOIprefix\doi{10.1243/03093247V141011}.
\bibitem[{Lorenz(2012)}]{lorenz2012mikrostruktur}
\bibinfo{author}{Lorenz\xfnm[ H.]}.
\newblock \bibinfo{title}{Mikrostruktur-basierte {M}odellierung des
  mechanischen {V}erhaltens {F\"u}llstoff-verst{\"a}rkter Elastomere}.
\newblock Ph.D. thesis; Technical Universsity of Dresden;
  \bibinfo{address}{Dresden}; \bibinfo{year}{2012}.
\bibitem[{Marckmann and Verron(2006)}]{marckmann2006}
\bibinfo{author}{Marckmann\xfnm[ G.]}, \bibinfo{author}{Verron\xfnm[ E.]}.
\newblock \bibinfo{title}{Comparison of hyperelastic models for rubber-like
  materials}.
\newblock \bibinfo{journal}{Rubber Chemistry and Technology}
  \bibinfo{year}{2006};\bibinfo{volume}{79}(\bibinfo{number}{5}):\bibinfo{pages}{835--858}.
\newblock \DOIprefix\doi{10.5254/1.3547969}.
\bibitem[{Martins et~al.(2006)Martins, Natal~Jorge and Ferreira}]{Martins2006}
\bibinfo{author}{Martins\xfnm[ P.A.L.S.]}, \bibinfo{author}{Natal~Jorge\xfnm[
  R.M.]}, \bibinfo{author}{Ferreira\xfnm[ A.J.M.]}.
\newblock \bibinfo{title}{A comparative study of several material models for
  prediction of hyperelastic properties: Application to silicone-rubber and
  soft tissues}.
\newblock \bibinfo{journal}{Strain}
  \bibinfo{year}{2006};\bibinfo{volume}{42}(\bibinfo{number}{3}):\bibinfo{pages}{135--147}.
\newblock \DOIprefix\doi{10.1111/j.1475-1305.2006.00257.x}.
\bibitem[{Miehe and G{\"o}ktepe(2005)}]{miehe2005micro}
\bibinfo{author}{Miehe\xfnm[ C.]}, \bibinfo{author}{G{\"o}ktepe\xfnm[ S.]}.
\newblock \bibinfo{title}{A micro--macro approach to rubber-like materials.
  {Part II}: the micro-sphere model of finite rubber viscoelasticity}.
\newblock \bibinfo{journal}{Journal of the Mechanics and Physics of Solids}
  \bibinfo{year}{2005};\bibinfo{volume}{53}(\bibinfo{number}{10}):\bibinfo{pages}{2231--2258}.
\bibitem[{Miehe et~al.(2004)Miehe, G{\"o}ktepe and Lulei}]{miehe2004micro}
\bibinfo{author}{Miehe\xfnm[ C.]}, \bibinfo{author}{G{\"o}ktepe\xfnm[ S.]},
  \bibinfo{author}{Lulei\xfnm[ F.]}.
\newblock \bibinfo{title}{A micro-macro approach to rubber-like materials.{Part
  I}: the non-affine micro-sphere model of rubber elasticity}.
\newblock \bibinfo{journal}{Journal of the Mechanics and Physics of Solids}
  \bibinfo{year}{2004};\bibinfo{volume}{52}(\bibinfo{number}{11}):\bibinfo{pages}{2617--2660}.
\bibitem[{Mooney(1940)}]{mooney1940theory}
\bibinfo{author}{Mooney\xfnm[ M.]}.
\newblock \bibinfo{title}{A theory of large elastic deformation}.
\newblock \bibinfo{journal}{Journal of Applied Physics}
  \bibinfo{year}{1940};\bibinfo{volume}{11}(\bibinfo{number}{9}):\bibinfo{pages}{582--592}.
\newblock \DOIprefix\doi{10.1063/1.1712836}.
\bibitem[{Ogden(1972)}]{Ogden565}
\bibinfo{author}{Ogden\xfnm[ R.W.]}.
\newblock \bibinfo{title}{Large deformation isotropic elasticity - on the
  correlation of theory and experiment for incompressible rubberlike solids}.
\newblock \bibinfo{journal}{Proceedings of the Royal Society of London A:
  Mathematical, Physical and Engineering Sciences}
  \bibinfo{year}{1972};\bibinfo{volume}{326}(\bibinfo{number}{1567}):\bibinfo{pages}{565--584}.
\newblock \DOIprefix\doi{10.1098/rspa.1972.0026}.
\bibitem[{Rebouah and Chagnon(2014)}]{rebouah2014permanent}
\bibinfo{author}{Rebouah\xfnm[ M.]}, \bibinfo{author}{Chagnon\xfnm[ G.]}.
\newblock \bibinfo{title}{Permanent set and stress-softening constitutive
  equation applied to rubber-like materials and soft tissues}.
\newblock \bibinfo{journal}{Acta Mechanica}
  \bibinfo{year}{2014};\bibinfo{volume}{225}(\bibinfo{number}{6}):\bibinfo{pages}{1685--1698}.
\bibitem[{Rivlin(1948)}]{Rivlin379}
\bibinfo{author}{Rivlin\xfnm[ R.S.]}.
\newblock \bibinfo{title}{Large elastic deformations of isotropic materials.
  iv. further developments of the general theory}.
\newblock \bibinfo{journal}{Philosophical Transactions of the Royal Society of
  London A: Mathematical, Physical and Engineering Sciences}
  \bibinfo{year}{1948};\bibinfo{volume}{241}(\bibinfo{number}{835}):\bibinfo{pages}{379--397}.
\newblock \DOIprefix\doi{10.1098/rsta.1948.0024}.
\bibitem[{Smeulders and Govindjee(1999)}]{smeulders1999phenomenological}
\bibinfo{author}{Smeulders\xfnm[ S.B.]}, \bibinfo{author}{Govindjee\xfnm[ S.]}.
\newblock \bibinfo{title}{A phenomenological model of an elastomer with an
  evolving molecular weight distribution}.
\newblock \bibinfo{journal}{Journal of Rheology}
  \bibinfo{year}{1999};\bibinfo{volume}{43}(\bibinfo{number}{2}):\bibinfo{pages}{393--414}.
\bibitem[{Steinmann et~al.(2012)Steinmann, Hossain and Possart}]{Steinmann2012}
\bibinfo{author}{Steinmann\xfnm[ P.]}, \bibinfo{author}{Hossain\xfnm[ M.]},
  \bibinfo{author}{Possart\xfnm[ G.]}.
\newblock \bibinfo{title}{Hyperelastic models for rubber-like materials:
  consistent tangent operators and suitability for treloar's data}.
\newblock \bibinfo{journal}{Archive of Applied Mechanics}
  \bibinfo{year}{2012};\bibinfo{volume}{82}(\bibinfo{number}{9}):\bibinfo{pages}{1183--1217}.
\newblock \DOIprefix\doi{10.1007/s00419-012-0610-z}.
\bibitem[{Swanson et~al.(1985)Swanson, Christensen and Ensign}]{SWANSON198581}
\bibinfo{author}{Swanson\xfnm[ S.]}, \bibinfo{author}{Christensen\xfnm[ L.]},
  \bibinfo{author}{Ensign\xfnm[ M.]}.
\newblock \bibinfo{title}{Large deformation finite element calculations for
  slightly compressible hyperelastic materials}.
\newblock \bibinfo{journal}{Computers \& Structures}
  \bibinfo{year}{1985};\bibinfo{volume}{21}(\bibinfo{number}{1}):\bibinfo{pages}{81--88}.
\newblock \DOIprefix\doi{10.1016/0045-7949(85)90231-7}.
\bibitem[{Thomas(1955)}]{thomas1955}
\bibinfo{author}{Thomas\xfnm[ A.G.]}.
\newblock \bibinfo{title}{The departures from the statistical theory of rubber
  elasticity}.
\newblock \bibinfo{journal}{Transactions of the Faraday Society}
  \bibinfo{year}{1955};\bibinfo{volume}{51}:\bibinfo{pages}{569--582}.
\newblock \DOIprefix\doi{10.1039/TF9555100569}.
\bibitem[{Treloar(1954)}]{treloar1954photoelastic}
\bibinfo{author}{Treloar\xfnm[ L.R.G.]}.
\newblock \bibinfo{title}{The photoelastic properties of short-chain molecular
  networks}.
\newblock \bibinfo{journal}{Transactions of the Faraday Society}
  \bibinfo{year}{1954};\bibinfo{volume}{50}:\bibinfo{pages}{881--896}.
\bibitem[{Vahapoğlu and Karadeniz(2006)}]{vahapoglu2006}
\bibinfo{author}{Vahapoğlu\xfnm[ V.]}, \bibinfo{author}{Karadeniz\xfnm[ S.]}.
\newblock \bibinfo{title}{Constitutive equations for isotropic rubber-like
  materials using phenomenological approach: A bibliography (1930-–2003)}.
\newblock \bibinfo{journal}{Rubber Chemistry and Technology}
  \bibinfo{year}{2006};\bibinfo{volume}{79}(\bibinfo{number}{3}):\bibinfo{pages}{489--499}.
\newblock \DOIprefix\doi{10.5254/1.3547947}.
\bibitem[{Wu and van~der Giessen(1993)}]{wu1993improved}
\bibinfo{author}{Wu\xfnm[ P.D.]}, \bibinfo{author}{van~der Giessen\xfnm[ E.]}.
\newblock \bibinfo{title}{On improved network models for rubber elasticity and
  their applications to orientation hardening in glassy polymers}.
\newblock \bibinfo{journal}{Journal of the Mechanics and Physics of Solids}
  \bibinfo{year}{1993};\bibinfo{volume}{41}(\bibinfo{number}{3}):\bibinfo{pages}{427--456}.
\bibitem[{Yeoh(1990)}]{yeoh1990}
\bibinfo{author}{Yeoh\xfnm[ O.H.]}.
\newblock \bibinfo{title}{Characterization of elastic properties of
  carbon-black-filled rubber vulcanizates}.
\newblock \bibinfo{journal}{Rubber Chemistry and Technology}
  \bibinfo{year}{1990};\bibinfo{volume}{63}(\bibinfo{number}{5}):\bibinfo{pages}{792--805}.
\newblock \DOIprefix\doi{10.5254/1.3538289}.
\bibitem[{Yeoh(1993)}]{yeoh1993}
\bibinfo{author}{Yeoh\xfnm[ O.H.]}.
\newblock \bibinfo{title}{Some forms of the strain energy function for rubber}.
\newblock \bibinfo{journal}{Rubber Chemistry and Technology}
  \bibinfo{year}{1993};\bibinfo{volume}{66}(\bibinfo{number}{5}):\bibinfo{pages}{754--771}.
\newblock \DOIprefix\doi{10.5254/1.3538343}.
\bibitem[{Yeoh and Fleming(1997)}]{yeoh1997}
\bibinfo{author}{Yeoh\xfnm[ O.H.]}, \bibinfo{author}{Fleming\xfnm[ P.D.]}.
\newblock \bibinfo{title}{A new attempt to reconcile the statistical and
  phenomenological theories of rubber elasticity}.
\newblock \bibinfo{journal}{Journal of Polymer Science Part B: Polymer Physics}
  \bibinfo{year}{1997};\bibinfo{volume}{35}(\bibinfo{number}{12}):\bibinfo{pages}{1919--1931}.
\newblock
  \DOIprefix\doi{10.1002/(SICI)1099-0488(19970915)35:12<1919::AID-POLB7>3.0.CO;2-K}.

\end{thebibliography}

\section*{Vitae}

Olaf Hesebeck studied physics at G{\"o}ttingen University and made his PhD in mechanical engineering at Karlsruhe University with a thesis on continuum damage mechanics. Since 2000 he works at the Fraunhofer Institute for Manu\-facturing Technology and Advanced Materials in Bremen, Germany, which is one of the major European research institutes for adhesive joining technology. Main focus of his research work is the modelling and mechanical testing of adhesive joints.

Andreas Wulf studied mechanical engineering at Hannover University. From 1991 to 1999 he worked at the Fraunhofer Institute for Manufacturing Techno\-logy and Advanced Materials in Bremen, Germany. Between 1999 and 2012 he worked at BASF Polyurethanes GmbH in Lemf{\"o}rde, Germany focused on microcellular polyurethane elastomers. Since 2013 he again works at the Fraunhofer Institute for Manufacturing Technology and Advanced Materials. Main focus of his research work is modelling and testing of adhesive joints with low-modulus elastic adhesives.

\end{document}